\documentclass[prd,showpacs,amsmath,amssymb,superscriptaddress,preprintnumbers,nofootinbib,showkeys]{revtex4}

\begin{document}

\title{Einstein-aether theory with a Maxwell field: General formalism}

\author{Alexander B. Balakin}
\email{Alexander.Balakin@kpfu.ru} \affiliation{Department of
General Relativity and Gravitation, Institute of Physics, Kazan
Federal University, Kremlevskaya str. 18, Kazan 420008,
Russia}\footnote{Corresponding author}
\author{Jos\'e P. S. Lemos}
\email{joselemos@ist.utl.pt}
\affiliation{Centro Multidisciplinar de Astrof\'{\i}sica-CENTRA
Departamento de F\'{\i}sica,
Instituto Superior T\'ecnico-IST,\\
Universidade de Lisboa-UL,
Avenida Rovisco Pais 1, 1049-001 Lisboa, Portugal}

\date{\today}
%
\begin{abstract}
We extend the Einstein-aether theory to include the Maxwell field in a
nontrivial manner by taking into account its interaction with the
time-like unit vector field characterizing the aether.  We also
include a generic matter term.  We present a model with a Lagrangian
that includes cross-terms linear and quadratic in the Maxwell tensor,
linear and quadratic in the covariant derivative of the aether
velocity four-vector, linear in its second covariant derivative and in
the Riemann tensor.  We decompose these terms with respect to the
irreducible parts of the covariant derivative of the aether velocity,
namely, the acceleration four-vector, the shear and vorticity tensors,
and the expansion scalar. Furthermore, we discuss the influence of an
aether non-uniform motion on the polarization and magnetization of the
matter in such an aether environment, as well as on its dielectric and
magnetic properties. The total self-consistent system of equations for
the electromagnetic and the gravitational fields, and the dynamic
equations for the unit vector aether field are obtained. Possible
applications of this system are discussed.  Based on the principles of
effective field theories, we display in an appendix all the terms up
to fourth order in derivative operators that can be considered in a
Lagrangian that includes the metric, the electromagnetic and the
aether fields.

\end{abstract}
\pacs{04.20.-q, 04.40.-b, 04.40.Nr, 04.50.Kd}
\keywords{Alternative theories of gravity, Einstein-aether theory,
unit vector field, electrodynamics of aether}
\maketitle

\section{Introduction}

The Einstein-aether theory is an alternative theory of gravity in
which, in addition to the spacetime metric, there is a non-vanishing
everywhere dynamic time-like unit vector field $U^i$ characterizing
the velocity of a substratum, the aether (see, e.g.,
\cite{J1,J2,J3,J41,J42,J5,J6,J7,bbm,J8} for reviews and references).

The Einstein-aether theory has attracted some attention for at least
five main motives. First, it is a pure field theory, i.e., a
vector-tensor theory of gravity \cite{CW} (see also
\cite{Vec0,Vec1}), admitting a rigorous formulation based on a
Lagrange formalism. Second, it realizes the idea of a preferred frame
of reference (see, e.g., \cite{N1,N2,N3}) associated with a world-line
congruence for which the corresponding time-like four-vector $U^i$ is
the tangent vector. Third, this time-like unit vector field $U^i$ can
be interpreted as a velocity four-vector of some medium-like
substratum (aether, vacuum, dark fluid, and so on), bringing into
consideration well-verified ideas and well-elaborated methods from the
relativistic theory of non-uniformly moving continuous media and their
interactions with other fields, such as the electromagnetic field
\cite{Mo,ME,LL,HehlObukhov}.  Fourth, the Einstein-aether theory is
also a specific realization of the idea of dynamic self-interaction of
complex systems moving with a spacetime dependent macroscopic
velocity.  Irregularities of the macroscopic motion are known to
influence the internal structure of complex systems and evolution of
their subsystems (see, e.g., \cite{LL}).  When we deal with an
accelerated expansion of the universe this dynamic self-interaction
can produce the same cosmological effects as the ones prescribed to
the dark energy, as it was shown in \cite{BD09}. Fifth, the
Einstein-aether theory, since it has a preferred unit vector field, is
characterized by a violation of Lorentz invariance. Theories admitting
Lorentz invariance violation are widely discussed in the literature.
In this instance it is supposed that the quantum gravity scale
provides a cutoff for the spacetime continuum, breaking thus, at some
stage, Lorentz invariance \cite{LIV31,LIV2,LIV3,LIV4,LIV1}. Various
constraints for the scale of the breaking coming from astrophysical
and cosmological observations have been obtained (see, e.g.,
\cite{LIV31,LIV2,LIV3,LIV4,LIV1} for reviews, details and references).

The Einstein-aether theory in the version elaborated by Jacobson and
colleagues
\cite{J1,J2,J3,J41,J42,J5,J6,J7,bbm,J8} is mainly motivated on the
grounds that Lorentz symmetry is broken at some scale, and the
decomposition of the Lagrangian used in their Einstein-aether theory
and in its extensions, can be naturally interpreted in terms of a low
energy effective theory \cite{EFT1,EFT2,EFT3,EFT4,EFT5}.  The
rationale and the heuristics for establishing such a low energy
effective theory have a parallel to other effective field theories,
notably in the effective quantum field theory generated by Goldstone
bosons of a chiral theory with a spontaneously broken symmetry
\cite{L1,L2}.

The Einstein-aether theory \cite{J1,J2,J3,J41,J42,J5,J6,J7} contains
four coupling parameters, which are to be estimated in gravitational
tests.  Indeed, the theory is within the sphere of analysis of three
parameterized formalisms used to test gravity theories. These are the
PPN (parameterized Post-Newtonian) \cite{CW} (see also \cite{J6}), the
PPE (parameterized Post-Einsteinian) \cite{3} and the PPF
(parameterized Post-Friedmann) \cite{2} formalisms. The classical PPN
formalism is focused on tests in the solar system, whereas the PPE and
PPF formalisms deal with strong gravitation.  Constraints on the
Einstein-aether theory have been already performed \cite{wave1,wave2}.

It is of course of interest to extend the Einstein-aether theory to
include other fields. The next most ubiquitous field is the
electromagnetic field.  When Lorentz symmetry is broken, and somehow a
preferred unit vector field pops up leading to an Einstein-aether
theory, there are certainly other fields around which would interact
with the metric and the aether. One of these fields could be the
electromagnetic field as we know it now or some version of a 2-form
field appropriate to the primordial universe.  In this connection, an
Einstein-scalar-aether theory, as an extension to the Einstein-aether
theory, has been proposed in \cite{solbar} to examine possible Lorentz
invariance violations in an inflationary period.
There is also a model, called the
bumblebee model, which introduces a Lorentz violating vector
field $B_k$ subject to some potential, and explores
the dynamics of its evolution that can be put
interacted with the
metric field
(see, e.g., \cite{bumblebee1,bumblebee2,bumblebee3}).
In this perspective,
these theories, namely the Einstein-Maxwell-aether theory we propose
here, the Einstein-scalar-aether proposed in \cite{solbar},
and the bumblebee theory \cite{bumblebee1,bumblebee2,bumblebee3},
should
be considered as effective field theories generated perhaps from a
fundamental quantum gravity operating at the Planck scale,
or from some other quantum field theory at a different scale, that has
some of its symmetries spontaneously broken at some stage (see
\cite{EFT1,EFT2,EFT3,EFT4,EFT5} and \cite{L1,L2}).
For instance, one might assume that
the appropriate quantum fundamental theory,
for which the
Einstein-Maxwell-aether
represents an effective field theory,
is a quantum version of theories
with effective metrics, i.e., theories
that associate and unify through effective metrics, optical,
color, and color-acoustic phenomena (see, e.g.,
\cite{OM1,OM2,CM1,CM2,CM3} for details and references). In these
settings,
the classical optical metric is composed of a spacetime metric
$g_{ik}$ and unit vector field $U_i$, and the metric has the form
$g^{*}_{ik}{=}Ag_{ik}+B U_i U_k$ for some appropriate scalar functions
$A$ and $B$. Photon propagation in a  medium that has
a velocity four-vector $U^i$ in the given spacetime
metric $g_{ik}$ is  equivalent to photon
propagation
along a geodesic line in an effective spacetime with
optical metric $g^{*}_{ik}$. The
corresponding version of quantized theory is not yet elaborated,
but the effective metric approach is rather
promising at a classical level.

In order to further justify extending the Einstein-aether theory to include
the electromagnetic field
at a classical level and take into account its interaction with
the gravitational field and the dynamic unit vector field, one can
consider several settings in which this interaction is important. The
first setting is connected with cosmology. The accelerated expansion
of the universe makes the aether motion irregular, and so the
interaction of electromagnetic waves with a non-uniformly moving
aether can change some fine details of the standard history of the
relic photons. A refined structure of the relic photon distribution in
the framework of the Einstein-Maxwell-aether theory could be tested
using WMAP data for the cosmic microwave background radiation. The
second situation in which the interaction is important appears in the
context of objects with strong gravitation, such as black holes,
wormholes and neutron stars.  Much of the information we may have from
these objects is from electromagnetic radiation coming from their
vicinity.  The interaction of this electromagnetic radiation with a
deformed aether in a strong gravitational field will induce new
dynamo-optical effects, which could be tested using observational
data.  A third situation that might be of relevance in this study is
related to gravitational waves.  The Einstein-Maxwell-aether theory
should also break Lorentz invariance, since the dynamic unit vector
field (velocity four-vector of the aether motion) remains one of the
basic elements of the extended theory.  The Einstein-Maxwell-aether
theory, as a theory with a preferred frame of reference, is expected
to predict new forms for gravitational wave propagation and consequent
detection (here our expectations are connected with generalizations of
the results obtained in the works \cite{wave1,wave2}).

The Einstein-Maxwell-aether model should be
experimentally verified. When one deals with an effective field
theory coming from some Lorentz symmetry violation
process, there are
constraints coming from astrophysical and cosmological
observations (see, e.g., the  data published in \cite{LIV3}).
Part of these data could be used to test the
Einstein-Maxwell-aether theory also. For instance, an analysis of
the gamma-ray burst observations (see, e.g., the results of
the
Fermi Large Area Telescope \cite{Fermi}) has shown that the
method known as modified photon dispersion relation, gives
an
estimation for the quantum gravity energy scale
$E_{(\rm QG)}$. The results are
 $E_{(\rm QG,1)} > 7.6 \times E_{(\rm Planck)}$ and
$E_{(\rm QG,2)} > 1.3 \times 10^{11} {\rm GeV}$ for the linear
($E_{(\rm QG,1)}$)
and
quadratic ($E_{(\rm QG,2)}$)
leading order terms, in the decomposition
of the dispersion function $f(k)=(\omega^2{-}k^2c^2)$ with
respect to the power law terms $\left[\frac{(kc)^{2{+}n}}{E_{(\rm
QG,n)}}\right]$, $n=1,2$, related to corrections induced by a Lorentz
symmetry violation.
The Einstein-Maxwell-aether
theory predicts effects of
electromagnetic polarization rotation which, in principle
could be detected in
X-ray and
$\gamma$-ray data.

Our goal is thus to extend the Einstein-aether theory by including a
Maxwell electromagnetic coupling to the gravitational field, to the
aether time-like unit vector field, and to other matter fields, in
short to study the Einstein-Maxwell-aether theory. For this purpose we
insert into the Einstein-aether Lagrangian all possible cross-terms,
which, on the one hand, are linear and quadratic in the Maxwell tensor
and, on the other hand, linear, quadratic and of the second order in
the covariant derivative of the aether velocity four-vector, as well
as linear in the curvature tensor and its convolutions.  In order to
classify, in a phenomenological way, the coupling constants appearing
in this Einstein-Maxwell-aether theory, we use the decomposition of
this covariant derivative of the aether velocity four-vector with
respect to its irreducible parts, namely, the acceleration
four-vector, the shear and vorticity tensors, and the expansion
scalar. This classification includes the set of independent coupling
constants related, first, to the effects of induced
polarization-magnetization of the matter in the moving aether, and
second, to the phenomena associated with optical activity,
birefringence, and so on. We should stress that these phenomenological
coupling constants can be, in principle, estimated in electromagnetic
and gravitational tests, thus extending the schemes of PPE and PPF
formalisms. It is then possible to find solutions of our
Einstein-Maxwell-aether theory. We give some hints how the symmetries
of the theory
can be used in some spacetime models, but we do not attempt
to find exact or numerical solutions.
The Einstein-scalar-aether theory proposed in
\cite{solbar} has interesting inflationary solutions,
and displays the possibilities offered by the
Einstein-Maxwell-aether theory we are proposing here.
The discussion of observational
effects of Lorentz invariance violation in our extended
Einstein-aether theory, although of importance,
is out of framework of this paper.

The paper is organized as follows. In Section~\ref{principles} we 
review the basic elements of the Einstein-aether theory.  In
Section~\ref{ematheory} based on our action functional for the
extended Einstein-Maxwell-aether theory we derive the 
equations for the electromagnetic, aether time-like unit vector and
gravitational fields. We also compare our theory 
with the bumblebee model and the work of Kostelecky and Mewes.
In Section~\ref{coupling} we decompose the
polarization and magnetization four-vectors and permittivity tensors
with respect to the acceleration four-vector, the shear and vorticity
tensors, and the expansion scalar, and classify the corresponding
coupling constants. We also discuss the Einstein-Maxwell-aether theory
for spacetimes of three types, namely, homogeneous isotropic Friedmann
cosmological models, static models with spherical symmetry, and
plane-wave models. In Section~\ref{conc} we draw some conclusions.
The Appendix~\ref{appA} displays all the terms up to fourth order in
derivative operators that can be considered in a Lagrangian that
includes the metric, the electromagnetic and the aether fields.  The
Appendix~\ref{App:AppendixB} contains further analysis and detailed
representation of the phenomenological tensors introduced in the
theory.

\section{Einstein-aether theory}
\label{principles}

\noindent
Einstein's theory is constructed from a Lagrangian
with the metric $g_{ab}$ and its two or fewer derivatives.
In order to construct an Einstein-aether theory
with a dynamic unit time-like vector field $U^a$
associated to the four-velocity of a background aether
one can think in adding to the Einstein theory
terms involving $U^a$
and its two or fewer derivatives \cite{J1}.

The Einstein-aether theory with a dynamic unit time-like vector field
associated to the four-velocity of a background aether can be
constructed using the following action functional \cite{J1}
\begin{equation}
S_{({\rm EA})} = \int d^4 x \sqrt{{-}g} \ \left\{
\frac{1}{2\kappa} \left[R{+}2\Lambda {+} \lambda \left(g_{mn}U^m
U^n {-}1 \right) {+} K^{abmn}  \ \nabla_a U_m  \ \nabla_b U_n
\right]
 {+} L_{({\rm m})} \right\} \,.
\label{1}
\end{equation}
The determinant of the metric $g = {\rm det}(g_{ik})$, the Ricci
scalar $R$, the cosmological constant $\Lambda$, and the matter
Lagrangian $L_{({\rm m})}$ are standard elements of the
Einstein-Hilbert action.
The new elements appearing in
Eq.~(\ref{1}) are the terms involving the vector field $U^i$. The
first such term $\lambda \left(g_{mn}U^m U^n {-}1 \right) $ ensures
that the $U^i$ is normalized to one, and the second term $K^{abmn} \
\nabla_a U_m \ \nabla_b U_n $ is quadratic in the covariant derivative
$\nabla_a U_m $ of the vector field $U^i$, with
$K^{abmn}$ a tensor field constructed
using the metric tensor $g^{ij}$ and the
velocity four-vector $U^k$ only,
\begin{equation}
K^{abmn} {=} C_1 g^{ab} g^{mn} {+} C_2 g^{am}g^{bn}
{+} C_3 g^{an}g^{bm} {+} C_4 U^{a} U^{b}g^{mn}
\label{2}
\end{equation}
and
where $C_1$, $C_2$, $C_3$ and $C_4$ are the Jacobson constants
that can be found from experiments or from some fundamental
theory.

The aether dynamic equations and the gravitational field
equations are found by varying the action (\ref{1}) with
respect to the vector field $U^i$ and the gravitational
field $g^{ij}$, respectively.
This procedure is well documented. Nevertheless, we recall
some of its
details, to make clearer the development we propose,
namely, to include the Maxwellian electromagnetic field,
and to introduce the nomenclature and
the standard definitions and relations.
Let us find these equations.

First, the term $\lambda$ is a Lagrange multiplier.
The variation of the action (\ref{1}) with respect to
$\lambda$ yields the equation
\begin{equation}
g_{mn}U^m U^n = 1 \,,
\label{21}
\end{equation}
which is the normalization condition of the time-like
vector field $U^k$.
Then, variation of the functional (\ref{1}) with respect to
$U^i$ yields that $U^i$ itself satisfies the equation
\begin{equation}
\nabla_m {\cal J}^{mn}_{({\rm A})}
- I^n_{({\rm A})} - \kappa I^n_{({\rm m})} = \lambda \ U^n  \,.
\label{0A1}
\end{equation}
Here we are using the standard definition
\begin{equation}
{\cal J}^{mn}_{({\rm A})} =  K^{lmsn} \nabla_l U_s \,,
\label{0A31}
\end{equation}
and have introduced two four-vectors
\begin{equation}
I^n_{({\rm A})} =  \frac12 \nabla_l U_s \nabla_m U_j \
 \frac{\delta K^{lsmj}}{\delta U_n}  =  C_4  DU_m \nabla^n U^m
\,,
\label{0A4A}
\end{equation}
and
\begin{equation}
I^n_{({\rm m})} =  \frac{\delta L_{({\rm m})}}{\delta U_n}
\,,
\label{0A4m}
\end{equation}
where $D$ appearing in Eq.~(\ref{0A4A}) is defined as $ D\equiv
U^i\,\nabla_i$. In comparison with \cite{J1} a new contribution,
$-\kappa I^n_{({\rm m})} =-\kappa \frac{\delta L_{({\rm
m})}}{\delta U_n}$ has appeared in Eq.~(\ref{0A1}), since now we
are assuming that the unit vector field is coupled to the matter.
The Lagrange multiplier has the following form
\begin{equation}
\lambda =  \lambda_{({\rm A})} + \kappa \lambda_{({\rm m})}
\,
\label{0A30}
\end{equation}
with
\begin{equation}
\lambda_{({\rm A})} = U_n \left[\nabla_m {\cal J}^{mn}_{({\rm A})}
- I^n_{({\rm A})}\right]  \,,  \label{0A309}
\end{equation}
and
\begin{equation}
\lambda_{({\rm m})} = - U_n \frac{\delta L_{({\rm m})}}{\delta U_n} \,.
\label{0A301}
\end{equation}
Using the projector $\Delta_n^j$
of tensors into the space orthogonal to
$U^i$,
$\Delta_n^j \equiv \delta_n^j {-}U_n U^j$, Eq.~(\ref{0A1})
can be rewritten as
\begin{equation}
\Delta_n^j \left[\nabla_m {\cal J}^{mn}_{({\rm A})} -
I^n_{({\rm A})} - \kappa I^n_{({\rm m})}\right] = 0 \,.
\label{0A19}
\end{equation}
The variation of the action (\ref{1}) with respect to the metric
$g^{ik}$ yields the gravitational field equations in the form
\begin{equation}
R_{ik} - \frac{1}{2} R \ g_{ik} -  \Lambda g_{ik} = T^{({\rm U})}_{ik} +
\kappa T^{({\rm m})}_{ik} + \kappa T^{({\rm int})}_{ik}
\,. \label{0Ein1}
\end{equation}
The term $T^{({\rm U})}_{ik}$ describes
the stress-energy tensor associated with the self-gravitation
of the vector field $U^i$; it has the form:
$$
T^{({\rm U})}_{ik} = C_1\left(\nabla_mU_i \nabla^m U_k {-}
\nabla_i U_m \nabla_k U^m \right) {+} C_4 DU_i DU_k {+}
$$
\begin{equation}
{+}\frac12 g_{ik} {\cal J}^{am}_{({\rm A})} \nabla_a U_m {+}
\nabla^m \left[U_{(i}{\cal J}_{k)m}^{({\rm A})}\right] {-}
\nabla^m \left[{\cal J}_{m(i}^{({\rm A})}U_{k)} \right] {-}
\nabla_m \left[{\cal J}_{(ik)}^{({\rm A})} U^m\right] {+} U_iU_k
U_n \left[\nabla_m {\cal J}^{mn}_{({\rm A})} {-} I^n_{({\rm A})}
\right]\,, \label{0Ein5}
\end{equation}
where $p_{(i} q_{k)}{\equiv}\frac12 (p_iq_k{+}p_kq_i)$
denotes
symmetrization.  The tensor $T^{({\rm U})}_{ik}$ disappears when
the motion of the aether is uniform, i.e., $\nabla_iU_k{=}0$, and
$\frac{\delta L_{({\rm m})}}{\delta U_n}{=}0$.  The term
\begin{equation}
T^{({\rm m})}_{ik} = - \frac{2}{\sqrt{{-}g}}\frac{
\delta}{\delta g^{ik}} \left[\sqrt{{-}g} L_{({\rm m})}\right]
\label{0Ein2}
\end{equation}
describes as usual the stress-energy tensor of the matter. The standard
algebraic decomposition of this tensor
\begin{equation}
T^{({\rm m})}_{ik} = W U_i U_k + I^{({\rm H})}_i U_k +
I^{({\rm H})}_k U_i + {\cal P}_{ik}
\label{0Ein3}
\end{equation}
introduces the energy density $W$, the heat-flux four-vector $I^{({\rm
H})}_i$, and the pressure tensor ${\cal P}_{ik}$, which are now determined
in the preferred frame of reference associated with the aether
velocity four-vector $U^i$, i.e.,
\begin{equation}
W = U^p T^{({\rm m})}_{pq} U^q \,, \quad I^{({\rm H})}_i =
\Delta^{p}_i T^{({\rm m})}_{pq} U^q \,, \quad {\cal P}_{ik} =
\Delta^{p}_i T^{({\rm m})}_{pq} \Delta^{q}_k = {-} P \Delta_{ik}
{+} \Pi_{ik} \,. \label{0T111}
\end{equation}
Here $P$ is the Pascal (isotropic) pressure, and $\Pi_{ik}$ is a
non-equilibrium pressure. The last term in
(\ref{0Ein1}) is due to the interaction between the
unit vector field and the matter
and it is described by
\begin{equation}
T^{({\rm int})}_{ik} = \lambda_{({\rm m})} U_i U_k =
- U_i U_k U_n \frac{\delta L_{({\rm m})}}{\delta U_n}
\,. \label{int}
\end{equation}
The compatibility conditions for the set of equations (\ref{0Ein1})
\begin{equation}
\nabla^k\left[ T^{({\rm U})}_{ik} +
\kappa T^{({\rm m})}_{ik} + \kappa T^{({\rm int})}_{ik}\right] = 0
\,, \label{compa1}
\end{equation}
involve all three quantities, thus showing that the stress-energy
tensor of the matter, $T^{({\rm m})}_{ik}$, is not  itself
a conserved
quantity because of the coupling of the aether to the matter.

Let us stress, that this interaction term describing a possible
coupling between the matter and the unit vector field has to be
postulated here, since we intend to generalize the Einstein-aether
theory by introduction a coupling between the electromagnetic field
and the unit vector field. The interaction term guarantees consistency
of the whole theory.

\section{Einstein-Maxwell-aether theory}
\label{ematheory}

\subsection{The inclusion of the Maxwell field and of terms
up to fourth order in the derivatives in the Einstein-aether theory
and
the ansatz}
\label{inclusion}

We want to  include an electromagnetic gauge vector field $A_i$
to extend the Einstein-aether theory into an
Einstein-Maxwell-aether theory.
The corresponding
gauge invariant Maxwell tensor $F_{ik}$ is
\begin{equation}
F_{ik}{=}\nabla_i A_k {-}\nabla_k A_i\,.
\label{maxwellA1}
\end{equation}

According to the principles of effective field theories (see,
e.g., \cite{EFT1,EFT2,EFT3,EFT4,EFT5}) one can establish some
interrelations between the terms in the action functional and
differential operators of the first, second, and higher orders.
In the Appendix~\ref{appA} we give the complete set of terms
that could be included in the action if one selects
terms with derivatives up to the fourth order.
The theory that we propose does not require all these
terms, since we impose further requirements.

Indeed, to set our ansatz
we impose three requirements that our theory should satisfy:

\noindent
{(a)} The electrodynamics of the theory
must be linear in the Maxwell tensor $F_{ik}$ and of second order in
the partial derivatives of the electromagnetic potential four-vector
$A_i$.

\noindent {(b)} The dynamical equations for the unit vector field
$U^i$ are considered to be a set of quasilinear equations of
second order in their partial derivatives.
A note is in order:
According to the standard terminology in
mathematical physics quasilinear
means that the equations can be nonlinear in
the four-vector $U^i$ itself, nonlinear in the first covariant
derivative $\nabla_i U_k$, but the second partial
derivatives $\partial_{i} \partial_k U_s$ enters the equations
linearly with tensorial coefficients that can depend on $U^i$ and
$F_{mn}$, but can not contain $\nabla_i U_k$.

\noindent
{(c)} The equations for the gravitational
field are considered to be
equations of
second order in the partial derivatives of the metric
(similarly to the standard Einstein's and Einstein-aether
theories).

This ansatz (composed of requirements
(a), (b), and (c))
can be reformulated as the assumption that the discarded
terms have coefficients, phenomenologically introduced, that are small
enough in comparison with the non-discarded coupling constants.

In some sense, we have followed the rationale used for the
Einstein-aether theory, namely, that for regions where quantum gravity
is not anymore dominant and Lorentz symmetry is already broken by
those very quantum effects an Einstein-aether theory can naturally
appear \cite{J1}. Thus the energy scale for this model is below the
Planck scale.  It is clear, that
the quantum theory behind the Einstein-aether theory,
as an effective low energy theory, is still beyond grasp. In the same
way that one expects that general relativity is a low energy
phenomenon from some quantum gravity, or other fundamental theory, one
also expects that the Einstein-aether theory is a low energy
phenomenon of such a theory.  Which of the theories is the correct
one, experiments can in principle tell.  Now, as Solomon and Barrow
\cite{solbar} have recently proposed, such an Einstein-aether theory
might be operating at the inflationary energy scale. Thus an
Einstein-scalar-aether in this setting is in action from after the
Planck scale to the inflationary scale and after. Since a generic
2-form field,
like the Maxwell field, can appear at very high energy scales an
Einstein-Maxwell-aether could be in operation between these energies
and the Planck scale.  Or it could be in playing after inflation
decays and the matter fields, such as the Maxwell field, make their
appearance.

\subsection{Action functional}
\label{action}

Keeping in mind the restrictions discussed in the Sec.~\ref{inclusion},
we start with the following ansatz for the
action functional
\begin{equation}
S_{({\rm total})} = S_{({\rm EA})} + S_{({\rm EMA1})}+ S_{({\rm
EMA2})} + S_{({\rm EMA3})} + S_{({\rm NM1})} + S_{({\rm NM2})} \,.
\label{E1}
\end{equation}
This Einstein-Maxwell-aether action contains five new terms
relatively to the  Einstein-aether action of Eq.~(\ref{1}).
These new terms are
$S_{({\rm EMA1})}$, $S_{({\rm EMA2})}$, $S_{({\rm EMA3})}$,
$S_{({\rm NM1})}$ and $S_{({\rm NM2})}$.

The first additional term
\begin{equation}
S_{({\rm EMA1})} = \frac{1}{2} \int d^4 x \sqrt{{-}g} \
\left[A^{mnpq} \nabla_m U_n  {+} B^{mnlspq} (\nabla_l U_s)
(\nabla_m U_n) \ \right]F_{pq} \label{M1}
\end{equation}
is linear in the Maxwell tensor and does not contain the
Riemann tensor and its convolutions. It contains
the covariant derivative of the aether
velocity four-vector.

The second additional term
\begin{equation}
S_{({\rm EMA2})} = \frac{1}{4} \int d^4 x \sqrt{{-}g} \ \left\{
\frac{1}{\mu} \left[F_{ik}F^{ik} {+} 2(\varepsilon \mu
{-}1)F_{im}U^m F^{in}U_n \right] {+} \left[X^{mnikpq} {+}
Y^{mnlsikpq} (\nabla_l U_s) \right]
 F_{ik}F_{pq} \nabla_m U_n \right\}
\label{M20}
\end{equation}
is quadratic in $F_{mn}$ and does not contain
the Riemann tensor.  It contains
the covariant derivative of the aether
velocity four-vector.
The scalar quantities $\varepsilon$ and $\mu$ are the dielectric and
magnetic permittivities, respectively, of the matter immersed in the
aether. They are equal to one if one deals with
pure aether.

The third term
\begin{equation}
S_{({\rm EMA3})} = \frac{1}{4} \int d^4 x \sqrt{{-}g} \ F_{pq}
\left[2{\cal B}^{mlspq}  \nabla_{(m} \nabla_{l)} U_s + {\cal
Y}^{mlsikpq} F_{ik} \nabla_{(m} \nabla_{l)} U_s \right]
\label{EMA3}
\end{equation}
contains a second covariant derivative of the unit vector field.
The tensor quantities $A^{mnpq}$, $B^{mnlspq}$, ${\cal B}^{mlspq}$,
$X^{mnikpq}$, $Y^{mnlsikpq}$, and ${\cal Y}^{mnlsikpq}$,
describe electrodynamic
properties of the matter in the moving aether.
They are constructed using the
metric $g_{ik}$, the covariant constant Kronecker tensors
($\delta^i_k$, $\delta^{ik}_{ab}$ and higher order Kronecker tensors),
the Levi-Civita tensor $\epsilon^{ikab}$, and the unit vector field
$U^k$.

The first nonminimal term is
\begin{equation}
S_{({\rm NM1})} = \frac{1}{4} \int d^4 x \sqrt{{-}g}
\ {\cal R}^{ikmn} F_{ik}F_{mn}\,,
\label{M99}
\end{equation}
and does not contain the unit vector field $U^k$. Here,
\begin{equation}
{\cal R}^{ikmn} =  q_1 R g^{ikmn} + q_2 \Re^{ikmn} + q_3 R^{ikmn}
\label{sus1}
\end{equation}
is the nonminimal susceptibility tensor with
\begin{equation}
g^{ikmn} \equiv \frac{1}{2}(g^{im}g^{kn} {-} g^{in}g^{km}) \,,
\label{rrr}
\end{equation}
and
\begin{equation}
\Re^{ikmn} \equiv \frac{1}{2} (R^{im}g^{kn} {-} R^{in}g^{km} {+}
R^{kn}g^{im} {-} R^{km}g^{in}) \,. \label{rrrr}
\end{equation}
The constants $q_1$, $q_2$ and $q_3$ are nonminimal parameters
describing a linear coupling of the Maxwell tensor $F_{mn}$ with
the curvature (see, e.g., \cite{BL05} for details).

The second nonminimal term can be represented as
\begin{equation}
S_{({\rm NM2})} = \frac{1}{4} \int d^4 x \sqrt{{-}g} \left\{
S^{ikmnlspq} F_{ls} R_{ikmn} F_{pq} + 2 Q F_{pq} R^{pk} U_k U^q
\right\}\,, \label{M990}
\end{equation}
and includes the terms listed in
Eqs.~(\ref{list113}) and (\ref{list6}).

After displaying all the terms of importance, we can now display
the total action functional. It is given by
$$
S_{({\rm total})} = \int d^4 x \sqrt{{-}g} \ \left\{
\frac{1}{2\kappa} \left[R{+}2\Lambda {+} \lambda \left(g_{mn}U^m
U^n {-}1 \right) {+} K^{abmn}  \ (\nabla_a U_m)  \ (\nabla_b U_n)
\right]
 {+} L_{({\rm m})} + \right.
$$
$$
+ \frac{1}{2} \left[A^{mnpq}   {+}
 B^{mnlspq} (\nabla_l U_s) \right] F_{pq} \nabla_m U_n
+ \frac{1}{2} Q R^{pk} U_k U^q F_{pq}  +
$$
$$
{+} \frac{1}{4} \ {\cal R}^{ikmn} F_{ik}F_{mn} {+} \frac{1}{4}
S^{ikmnlspq} F_{ls} R_{ikmn} F_{pq} {+} \frac{1}{4} \left[2{\cal
B}^{mlspq} + {\cal Y}^{mlsikpq} F_{ik} \right] F_{pq} \nabla_{(m}
\nabla_{l)} U_s +
$$
\begin{equation}
\left.
{+}\frac{1}{4\mu} \left[F_{ik}F^{ik} {+} 2(\varepsilon \mu
{-}1)F_{im}U^m F^{in}U_n \right] {+} \frac{1}{4} \left[X^{mnikpq}
{+} Y^{mnlsikpq} (\nabla_l U_s) \right] (\nabla_m U_n)
F_{ik}F_{pq} \right\}\,. \label{M2}
\end{equation}
The coefficients involved into the decompositions of these
quantities can be interpreted as coupling constants (see
\cite{BL05} for the interpretation of the nonminimal coupling
constants $q_1$, $q_2$, $q_3$, and Section IV for the
interpretation of other coupling constants).

Given the action, Eq.~(\ref{M2}), we can now obtain the
electrodynamic equations, the aether dynamic equations and the
gravitational field equations, by variation of the corresponding
appropriate quantities.

\subsection{Electrodynamic equations}
\label{electroeq}

The electrodynamic equations can be
obtained by variation of the action (\ref{M2}) with
respect to the electromagnetic potential four-vector $A_i$, which
enters the Maxwell tensor $F_{ik}$ via $F_{ik}{=} \nabla_i A_k {-}
\nabla_k A_i$. The result of the variational procedure can be written
in the following standard form
\begin{equation}
\nabla_k H^{ik} = - 4\pi I^i \,,
\label{E2}
\end{equation}
where $H^{ik}$ is the excitation tensor linear in the Maxwell tensor
and given by
\begin{equation}
H^{ik} = {\cal H}^{ik} + C^{ikmn}F_{mn} \,,
\label{E3}
\end{equation}
where ${\cal H}^{ik}$, $ C^{ikmn}$, and $I^i$ have their own
physical meanings. The skew-symmetric tensor ${\cal H}^{ik}$ is
given by
\begin{equation}
{\cal H}^{ik} =  A^{mnik} \nabla_m U_n {+} B^{mnlsik} (\nabla_l
U_s)  (\nabla_m U_n) + {\cal B}^{mlsik}\nabla_{(m} \nabla_{l)} U_s
{+} Q U_m R^{m[i}U^{k]} \,, \label{E4}
\end{equation}
and describes the spontaneous polarization-magnetization of the
matter influenced by the moving aether. The tensor $C^{ikmn}$ is a
linear response tensor, and in turn, can be decomposed into four
terms, namely,
\begin{equation}
C^{ikmn} = C_{(0)}^{ikmn} {+} C_{({\rm D})}^{ikmn}
{+} C_{({\rm R})}^{ikmn} {+} C_{({\rm DD})}^{ikmn}\,.
\label{E5}
\end{equation}
The first term is given by
\begin{equation}
C_{(0)}^{ikmn}  = \frac{1}{2\mu} \left[ \left(g^{im}g^{kn} {-}
g^{in}g^{km} \right) {+} (\varepsilon \mu {-}1)\left(g^{im}U^k U^n
{-} g^{in} U^k U^m {+} g^{kn} U^i U^m {-} g^{km} U^i U^n \right)
\right] \,. \label{E6}
\end{equation}
It contains the four-vector $U^i$ but does not include
the covariant derivative $\nabla_m U_n$. The second term
is given by
\begin{equation}
C_{({\rm D})}^{ikmn} = X^{lsikmn} \nabla_l U_s + Y^{ablsikmn}
(\nabla_a U_b) (\nabla_l U_s)\,.
\label{E7}
\end{equation}
It
contains terms linear and quadratic in the covariant derivative
of the aether velocity four-vector $U^i$.
The third term is given by
\begin{equation}
C_{({\rm R})}^{ikmn} = {\cal R}^{ikmn} + S^{pqabikmn} R_{pqab}\,.
\label{E72}
\end{equation}
It is a
generalized nonminimal susceptibility tensor.
The fourth term is given by
\begin{equation}
C_{({\rm DD})}^{ikmn} = {\cal Y}^{jlsikmn}
\nabla_{(j} \nabla_{l)} U_s\,.
\label{E73}
\end{equation}
It relates to the linear response induced by a second covariant
derivative of the unit vector field $U^i$. The electric current
four-vector $I^i$, appearing in the right-hand side of
Eq.~(\ref{E2}), is defined as follows
\begin{equation}
I^i = \frac{1}{4\pi} \frac{\partial L_{({\rm m})}}{\partial A_i} \,.
\label{E8}
\end{equation}
As usual, we have to add the
Maxwell equation
\begin{equation}
\nabla_k F^{*ik} = 0 \,,
\label{E9}
\end{equation}
where the asterisk indicates dualization, i.e.,
\begin{equation}
F^{*ik} = \frac{1}{2} \epsilon^{ikmn} F_{mn} \,.
\label{E10}
\end{equation}
Here $\epsilon^{ikmn} {=} \frac{{\rm e}^{ikmn}}{\sqrt{{-}g}}$ is the
Levi-Civita tensor with ${\rm e}^{ikmn}$ being
the completely skew-symmetric symbol (${\rm e}^{0123}{=}1$).

Electrodynamics of
continuous media can be formulated in terms
of four-vectors representing
physical fields. These four-vector
fields are the electric field $E^i$, the magnetic
field ${\cal H}^i$, the electric excitation ${\cal D}^i$,
and the magnetic
excitation $B^i$ \cite{ME}: They are defined
in terms of $F^{ik}$ and $H^{ik}$ as,
\begin{equation}
E^i = F^{ik} U_k \,, \quad B^i = F^{*ik}U_k \,, \quad {\cal D}^i =
H^{ik} U_k \,, \quad {\cal H}^i = H^{*ik}U_k \,. \label{EHDB1}
\end{equation}
Other quantities useful to interpret the
phenomenological coupling constants
that appear in this formalism are the polarization vector,
$P^i$,
and the magnetization vector, $M^i$. These two
four-vectors
are defined as
\begin{equation}
P^i = {\cal D}^i - E^i \,, \quad M^i = {\cal H}^i - B^i \,,
\label{EHDB11}
\end{equation}
respectively.
The electrodynamic equations in terms of these
quantities contain covariant derivatives of the velocity
four-vector $U^i$.
The corresponding equations are written in
the Appendix~\ref{App:AppendixB}.

To be complete,
inverting Eq.~(\ref{EHDB1}), we find that
the tensors $F^{ik}$ and $H^{ik}$ can be written
in terms of $E^i$, $B^i$, ${\cal D}^i$, and  ${\cal H}^i$ as
\begin{equation}
F^{ik} = \delta^{ik}_{mn} E^m U^n - \epsilon^{ikmn} B_m U_n \,,
\quad  H^{ik} = \delta^{ik}_{mn} {\cal D}^m U^n - \epsilon^{ikmn}
{\cal H}_m U_n \,, \label{EHDB2}
\end{equation}
where $\delta^{ik}_{mn}$ and $ \epsilon^{ikmn}$
are the generalized Kronecker delta and the Levi-Civita
tensor, respectively.

\subsection{Aether dynamic equations}
\label{aethereqs}

The dynamic equations for the aether are obtained by the variation
of the action functional (\ref{M2}) with respect to the velocity
four-vector $U^i$. The variation procedure yields four equations
\begin{equation}
\nabla_m \left[{\cal J}^{mn}_{({\rm A})} {+} \kappa {\cal
J}^{mn}_{({\rm M})}\right] = I^n_{({\rm A})} {+} \kappa I^n_{({\rm
m})} {+} \kappa I^n_{({\rm M})} {+} \lambda \ U^n \,.
\label{A1}
\end{equation}
The Lagrangian multiplier is now
\begin{equation}
\lambda = \lambda_{({\rm A})} + \kappa \lambda_{({\rm m})} +
\kappa \lambda_{({\rm M})} \,,
\label{A30-1}
\end{equation}
where
$\lambda_{({\rm A})}$ and $ \lambda_{({\rm m})}$
are defined
in Eqs.~(\ref{0A309})-(\ref{0A301}),
respectively, and
\begin{equation}
\lambda_{({\rm M})}= U_n \left[\nabla_m {\cal J}^{mn}_{({\rm M})} -
I^n_{({\rm M})} \right]  \,.
\label{A30}
\end{equation}
Eliminating the Lagrange
multiplier $\lambda$ in Eq.~(\ref{A1}) one obtains the
following compact equation,
\begin{equation}
\Delta_n^s \left\{\nabla_m \left[{\cal J}^{mn}_{({\rm A})} {+} \kappa
{\cal J}^{mn}_{({\rm M})}\right] {-} \left[I^n_{({\rm A})} {+} \kappa
I^n_{({\rm m})} {+} \kappa I^n_{({\rm M})} \right] \right\} = 0 \,.
\label{A2}
\end{equation}
The quantities ${\cal J}^{mn}_{({\rm A})}$, $I^n_{({\rm A})}$, $I^n_{({\rm
m})}$, in  Eqs.~(\ref{A1})-(\ref{A2}), are defined
in Eqs.~(\ref{0A31})-(\ref{0A4m}), respectively.
The other two quantities that appear in
Eqs.~(\ref{A1})-(\ref{A2}) are defined as,
$$
{\cal J}^{mn}_{({\rm M})} =  \frac{1}{2} F_{pq}\left(A^{mnpq} + 2
B^{mnlspq} \nabla_l U_s \right) -
$$
\begin{equation}
- \frac14 \nabla_l\left[\left(2 {\cal B}^{(lm)npq} {+}{\cal
Y}^{(lm)nikpq} F_{ik}  \right)F_{pq}\right] + \frac{1}{4} F_{ik}
F_{pq}\left(X^{mnikpq} + 2 Y^{mnlsikpq} \nabla_l U_s \right) \,,
\label{A32}
\end{equation}
$$
I^n_{({\rm M})} = \left(\varepsilon {-}\frac{1}{\mu} \right)
F^{kn}F_{km} U^m  + \frac12 F_{pq} \nabla_l U_s \left(\frac{\delta
A^{lspq}}{\delta U_n} +
 \nabla_m U_j \frac{\delta B^{mjlspq}}{\delta U_n} \right)+
$$
$$
+ \frac14 F_{ik} F_{pq} \nabla_l U_s  \left(\frac{\delta
X^{lsikpq}}{\delta U_n} +
 \nabla_m U_j \frac{\delta Y^{mjlsikpq}}{\delta U_n} \right)
+ \frac14  R_{ikmj}F_{ls} F_{pq} \frac{\delta}{\delta U_n}
S^{ikmjlspq}+
$$
\begin{equation}
{+}\frac14 F_{pq}  \left[ 2 \frac{\delta}{\delta U_n} {\cal
B}^{mlspq} {+} F_{ik}\frac{\delta}{\delta U_n} {\cal Y}^{mlsikpq}
\right]\nabla_{(m} \nabla_{l)} U_s + \frac12 Q U^l
\left(R^{nm}F_{ml} - F^{nm}R_{ml} \right) \,. \label{A6}
\end{equation}

\subsection{The gravitational field equations}
\label{graveqs}

\subsubsection{The general equations}
\label{generalgraveqs}

The variation of the action functional (\ref{M2}) with respect to
the metric $g^{ik}$ yields

\begin{equation}
R_{ik} {-} \frac{1}{2} R \ g_{ik} {-}  \Lambda g_{ik} =
 T^{({\rm U})}_{ik} {+}\kappa \left[T^{({\rm m})}_{ik}
{+}T^{({\rm int})}_{ik}{+} T^{({\rm EM0})}_{ik}{+} T^{({\rm
EMA1})}_{ik} {+} T^{({\rm EMA2})}_{ik} {+} T^{({\rm EMA3})}_{ik}
{+} T^{({\rm NM1})}_{ik} {+} T^{({\rm NM2})}_{ik}\right] \,.
\label{Ein1}
\end{equation}
The terms $T^{({\rm U})}_{ik}$, $T^{({\rm m})}_{ik}$ and $T^{({\rm
int})}_{ik}$ are given by the formulas
(\ref{0Ein5}), (\ref{0Ein2})-(\ref{0T111}), and (\ref{int}),
respectively. Let us discuss in detail the new elements of this
decomposition.

\subsubsection{Stress-energy tensor of the electromagnetic field in
a uniformly moving aether}
\label{stressmaxwell}

The part of the stress-energy tensor of the electromagnetic field
indicated as $T^{({\rm EM0})}_{ik}$ is given by
\begin{equation}
T^{({\rm EM0})}_{ik} = \frac{1}{\mu} \left\{\left[\frac14
g_{ik}F_{mn}F^{mn}-F_{im}F_k^{\ m}\right] {+} (\varepsilon \mu {-}
1)U^p U^q\left[\left(\frac12 g_{ik}{-} U_iU_k\right) F^{m}_{\ \ p}
F_{mq} {-} F_{ip}F_{kq} \right] \right\} \,. \label{Ein6}
\end{equation}
In vacuum, $\varepsilon {=} \mu {=} 1$, $T^{({\rm EM0})}_{ik}$
gives the usual Maxwell term.
Clearly, the tensor (\ref{Ein6}) is symmetric and traceless,
i.e.,
\begin{equation}
T^{({\rm EM0})}_{ik} =  T^{({\rm EM0})}_{ki} \,, \quad
T^{({\rm EM0})}_{ik} g^{ik}=0 \,.
\label{Ein7}
\end{equation}
Other interesting quantities are connected with the energy density
scalar $W_{({\rm EM0})}$ and with the energy flux four-vector
$I^i_{({\rm EM0})}$ associated with this tensor. They are
related to $T^{({\rm EM0})}_{pq}$ and
defined as
\begin{equation}
W_{({\rm EM0})} \equiv  U^p T^{({\rm EM0})}_{pq} U^q =
{-}\frac12 \left(\varepsilon E^mE_m {+} \frac{1}{\mu} B^m B_m
\right) ,
\label{Ein8}
\end{equation}
\begin{equation}
I^i_{({\rm EM0})} \equiv \Delta^{ip} T^{({\rm EM0})}_{pq}U^q = -
\epsilon^{imns} U_s E_m {\cal H}_n \,.
\label{Ein81}
\end{equation}
Clearly, $W_{({\rm EM0})}$ coincides with the standard definition of
the energy density scalar in a spatially isotropic medium \cite{ME},
and $I^i_{({\rm EM0})}$ coincides with the Poynting vector. All these
properties allow us to identify this tensor $T^{({\rm EM0})}_{pq}$
with the stress-energy
tensor of the electromagnetic field in the Abraham version
\cite{Mo} (see also, \cite{ME}). Thus it is interesting to note
that, on the one hand, the tensor
$T^{({\rm EM0})}_{ik}$ is an effective stress-energy tensor, since it is
obtained by variation with respect to the metric \cite{B07}, on the
other hand, it coincides with the Abraham tensor, which appears from
an analysis of the balance equations in the electrodynamics of a
moving continuous medium \cite{Mo}.

\subsubsection{Stress-energy tensor associated with a spontaneous
polarization-magnetization of matter or vacuum induced by a
non-uniformly moving aether}
\label{stresspolar}

The quantity $T^{({\rm EMA1})}_{ik}$ appearing in  Eq.~(\ref{Ein1})
is given by
$$
T^{({\rm EMA1})}_{ik} = \frac12 g_{ik} F_{pq}
\left(A^{mnpq}+B^{mnlspq}\nabla_lU_s \right) \nabla_m U_n -
\frac12 \nabla_m \left\{F_{pq}U^m \left[A^{\ \ \ pq}_{(ik)}+ 2B^{\
\ \ lspq}_{(ik)}\nabla_lU_s \right]\right\}-
$$
$$
-\frac12 \nabla_m \left\{F_{pq} \left[U_{(i}A^{m \ pq}_{ \ k)}+
2U_{(i}B^{m \ lspq}_{\ k)}\nabla_lU_s \right]\right\}+ \frac12
\nabla_m \left\{F_{pq} \left[U_{(i}A^{\  mpq}_{k)}+2U_{(i} B^{ \
mlspq}_{k)}\nabla_lU_s \right]\right\}+
$$
$$
+ \frac12 U_i U_k U_n \left\{\nabla_m \left[F_{pq} \left(A^{mnpq}+
2B^{mnlspq}\nabla_lU_s \right)\right] - F_{pq}\nabla_l U_s
\left(\frac{\delta}{\delta U_n}A^{lspq}+ \nabla_{m}U_{j}
\frac{\delta}{\delta U_n}B^{mjlspq} \right)\right\} -
$$
\begin{equation}
- F_{pq} \nabla_mU^n \left(\frac{\delta}{\delta g^{ik}}
A^{m \ pq}_{\ \ n} + \nabla_lU^s\frac{\delta}{\delta g^{ik}}
B^{m \ l \ \ pq}_{\ \ n \ s} \right)
\,.
\label{Ein9}
\end{equation}
This stress-energy tensor is linear in the Maxwell
tensor $F_{pq}$ and therefore is
generated by a spontaneous
polarization-magnetization induced in the system due to
a non-uniformly moving aether.

\subsubsection{Stress-energy tensor of the electromagnetic field,
quadratic in the Maxwell tensor and quadratic in the vector field
covariant derivative}
\label{stressquadr}

The quantity $T^{({\rm EMA2})}_{ik}$ appearing in
Eq.~(\ref{Ein1}) is given by
$$
T^{({\rm EMA2})}_{ik} = \frac14 g_{ik} F_{ab}
F_{pq}\left(X^{mnabpq}{+} Y^{mnlsabpq}\nabla_lU_s \right) \nabla_m
U_n {-}
$$
$$
-\frac14 \nabla_m \left\{F_{ab}F_{pq}U^m \left[X^{\ \ \
abpq}_{(ik)}+2Y^{\ \ \ lsabpq}_{(ik)} \nabla_lU_s \right]\right\}-
\frac14 \nabla_m \left\{F_{ab}F_{pq} \left[U_{(i}X^{m \ abpq}_{\ \
k)}+2U_{(i}Y^{m \ lsabpq}_{\ \ k)}\nabla_lU_s \right]\right\}+
$$
$$
+\frac14 \nabla_m \left\{F_{ab}F_{pq} \left[U_{(i}X^{\ \
mabpq}_{k)}+2U_{(i}Y^{ \ \ mlsabpq}_{k)} \nabla_lU_s
\right]\right\}{+}\frac14 U_iU_kU_n \left\{\nabla_m
\left[F_{ab}F_{pq} \left(X^{mnabpq}{+}2Y^{mnlsabpq}\nabla_lU_s
\right)\right]{-} \right.
$$
\begin{equation}
\left. {-}F_{ab}F_{pq}\nabla_l U_s \left(\frac{\delta}{\delta
U_n}X^{lsabpq}{+}\nabla_{m}U_{j} \frac{\delta}{\delta
U_n}Y^{mjlsabpq} \right)\right\} {-} \frac12 F_{ab}F_{pq}
\nabla_mU^n \left(\frac{\delta}{ \delta g^{ik}} X^{m \ abpq}_{\ n}
{+} \nabla_lU^s\frac{\delta}{ \delta g^{ik}}Y^{m \ l \  abpq}_{\ \
n \ s} \right). \label{Ein10}
\end{equation}
This $T^{({\rm EMA2})}_{ik}$ is quadratic in the Maxwell tensor
and quadratic in the vector field covariant derivative. Note that,
in general, $T^{({\rm EMA1})}_{ik}$ and $T^{({\rm EMA2})}_{ik}$
are not traceless and are not conserved quantities.

\subsubsection{Stress-energy tensors of the electromagnetic
field in the aether environment,
nonminimally coupled to gravity}
\label{stressnonmin}

In Eq.~(\ref{Ein1}),
there are three nonminimal terms included in the
total stress-energy tensor, namely
$T^{({\rm NM1})}_{ik}$, $T^{({\rm
NM2})}_{ik}$, and $T^{({\rm EMA3})}_{ik}$.
We assume these terms have a negligible contribution
in the gravitational field
and do not discuss them further.

\subsection{Remarks}
\label{remarksbumbleandkosteleckymewes}

\subsubsection{Analogy with the bumblebee model}
\label{remarksbumble}

The set of master equations just derived points to an  analogy to the
bumblebee model \cite{bumblebee1,bumblebee2,bumblebee3}).  The
bumblebee model introduces a Lorentz violating vector field $B_k$ in
such a way that a scalar potential $V(B_kB^k)$ is inserted into the
Lagrangian.  There are also nonminimal terms of the form $R^{ik}B_i
B_k$ and other terms of this type.  The covariant derivative of the
bumblebee field, $\nabla_i B_k$, can be decomposed into the sum of
symmetric and skew-symmetric parts: $\nabla_i B_k {=} \nabla_{(i}
B_{k)} {+}\nabla_{[i} B_{k]}$. The skew-symmetric part of this tensor
can be (up to a factor 2), identified as an analog of the Maxwell
tensor in electrodynamics, i.e., $B_{ik} \equiv \nabla_i B_k
{-}\nabla_k B_i$.  In this sense the electromagnetic theory can be
extracted from this bumblebee model as a particular case.

Comparing the bumblebee model with the Einstein-Maxwell-aether theory
we propose here we mention three important facts. First, the
Einstein-Maxwell-aether model deals with two independent vector
fields, namely, the unit four-vector $U^i$ and the potential
four-vector $A_k$; on the other hand, the bumblebee model contains a
unique vector field $B_k$.  Second, the Einstein-Maxwell-aether
theory describes an aether field plus a Maxwell field, while with the
bumblebee model is suited to study or the aether, or the Maxwell
field.  Third, in our Lagrangian, we have considered terms in the
square of the covariant derivatives of the vector field $U_k$, terms
for the pure electromagnetic field, and then the cross-terms, which
contain both $\nabla_iU_k$ and $F_{ik}$. These cross-terms do not
appear in the original bumblebee model. The effects that come from the
coupling of a non-uniformly moving aether with the electromagnetic
field are specially interesting in the Einstein-Maxwell-aether theory.

\subsubsection{Comparison with the work of Kostelecky and Mewes}
\label{kostmewes1}

In a quite general theory,
from which  the standard model extension
is incorporated,
Kostelecky et.~al.~\cite{LIV31}
display the Lorentz violating terms containing tensor
coefficients.
It is of importance to give a comparison of the terms we use in our
action Eq.~(\ref{E1}) and subsequent equations with the terms given by
Kostelecky and Mewes \cite{LIV31}.
In particular, let us compare briefly the structure of
these Lorentz violating
coefficients of  \cite{LIV31},
which are formally similar to the ones introduced in our
work (see, e.g., our Eqs. (22), (23)),
and emphasize the novelty of our approach.

First, is it possible to extract all our tensor coefficients from the
terms $(k^{(d)}_{AF})_{\kappa}$ and $(k^{(d)}_{F})^{\kappa \lambda \mu
\nu}$ appearing in Eqs.~(9) and (10) of the paper \cite{LIV31} with
derivative operators of zero order?  The answer is negative, since our
tensors in Eqs.~(22) and (23) contain covariant derivatives of the
aether velocity unit four-vector $\nabla_i U_k$, namely, terms of
second order $(\nabla_i U_k) (\nabla_m U_n)$ and $\nabla_i \nabla_k
U_m$.  That is why in our case we deal, in fact, with tensors of
coefficients possessing five and six indices in those terms.  In
contrast, in \cite{LIV31} the authors use coefficients with three and
four indices. The physical interpretations of these coefficients
differ indeed, and the strategies to their experimental verification
also do not coincide.

Second, there is a similarity in that the Lagrangians in both works
are quadratic in the electromagnetic potential $A_i$.  In the paper
\cite{LIV31} one can find terms of the type $A_{\alpha_1}
\partial_{\alpha_{3}}...\partial_{\alpha_{d}} A_{\alpha_2}$ (see,
e.g., Eq.~(1) in \cite{LIV31}).  There are also gauge-invariant terms,
in which $A_{\alpha_1}$ is replaced by $F_{\mu \nu}$ (see, e.g.,
Eqs.~(8), (9), and (10) in \cite{LIV31}).  In the
Appendix~\ref{appA} we give
the terms cubic and of the fourth order in $F_{mn}$, but when we set
up our ansatz we discard these terms cubic and of the fourth order in
$F_{mn}$, and thus consider the terms linear and quadratic in
$F_{mn}$.

Third, in our work we excluded from the action functional all the
terms that contain derivatives of the Maxwell tensor $\nabla_k
F_{mn}$. These terms disappear either through an integration by parts,
or by using our ansatz. In the work \cite{LIV31} the derivatives of
the Maxwell tensor enter the basic decomposition as an essential
part. So in this particular item,
our classification can be considered as a subclassification
of the terms that are indicated as zero-order in the derivatives of
$F_{ik}$ in the work \cite{LIV31}.

\section{Classification of the coupling constants}
\label{coupling}

\subsection{Motivation}
\label{couplingmot}

The Einstein-Maxwell-aether theory proposed here contains several
coupling constants. Since the theory has been constructed in a
phenomenologically manner, these couplings have to be estimated and
determined experimentally.  In addition, underlying symmetries of the
models under study can be of help in finding some of the coupling
constants.

How many constants should we consider as key parameters for the
theory? In order to clarify this question, let us start with the known
discussion in the Einstein-aether theory about the number of
independent constants appearing in the formulation of the tensor
$K^{abmn}$. This tensor possesses the following symmetry of indices:
$K^{abmn}{=}K^{mnab}$.  Also, since $U^m \nabla_i U_m{=}0$, a number
of components of $K^{abmn}$ can be connected by the relations
$K^{abmn}U_m {=}0 {=}K^{abmn}U_n$.  Thus, there are, in principle,
$\frac12 \times 12 \times 13 {=} 78$ independent components for this
tensor. However, usually one deals with only four coupling constants
$C_1$, $C_2$, $C_3$, and $C_4$, related to $K^{abmn}$. We now recall
how this problem is solved in the Einstein-aether theory and then use
the same idea in the extended Einstein-Maxwell-aether theory.

\subsection{Decomposition of $\nabla_i U_k$ and interpretation of
Jacobson's coupling constants }
\label{decomposu}

The tensor $\Psi_{ik} \equiv \nabla_i U_k$ can be decomposed,
as usual, into a sum of
its irreducible parts,
namely, the acceleration four-vector $DU^{i}$,
the shear tensor $\sigma_{ik}$,
the vorticity tensor $\omega_{ik}$, and
the expansion scalar $\Theta$. The decomposition is given by
\begin{equation}
\Psi_{ik} = \nabla_i U_k = U_i DU_k + \sigma_{ik} + \omega_{ik} +
\frac{1}{3} \Delta_{ik} \Theta \,, \label{act3}
\end{equation}
where the basic quantities are defined as
$$
U_i\, DU_k \equiv U_i\, U^m \nabla_m U_k \,, \quad \sigma_{ik}
\equiv \frac{1}{2}\Delta_i^m \Delta_k^n \left(\nabla_m U_n {+}
\nabla_n U_m \right) {-} \frac{1}{3}\Delta_{ik} \Theta  \,,
$$
\begin{equation}
\omega_{ik} \equiv \frac{1}{2}\Delta_i^m \Delta_k^n \left(\nabla_m
U_n {-} \nabla_n U_m \right) \,, \quad \Theta \equiv \nabla_m U^m
\,, \quad D \equiv U^i \nabla_i \,. \label{act4}
\end{equation}
Now, when we construct the scalar $ K^{abmn}(\nabla_a
U_m)(\nabla_b U_n)$ using the unit vector field $U^m$ itself and
the geometric quantities $g^{ik}$, $\delta^i_k$,
$\delta^{ik}_{pq}$, $\epsilon^{ikmn}$, we find that there are 4
and only 4 non-vanishing independent second-order scalars
expressed in terms of $DU^i$, $\sigma_{mn}$, $\omega_{mn}$,
$\Theta$. These are $DU_k DU^k$, $\sigma_{ik} \sigma^{ik}$,
$\omega_{ik} \omega^{ik}$ and $\Theta^2$ \cite{J8}. Clearly, non-vanishing
cross-terms cannot be constructed. Thus, in these terms the scalar
$ K^{abmn}(\nabla_a U_m) (\nabla_b U_n)$ which appears in the
action functional given by Eq.~(\ref{M2}) should be represented as
\begin{equation}
K^{abmn}(\nabla_a U_m) (\nabla_b U_n) = (C_1 {+} C_4)DU_k DU^k {+}
(C_1 {+} C_3)\sigma_{ik} \sigma^{ik} {+}  (C_1 {-} C_3)\omega_{ik}
\omega^{ik} {+} \frac13 \left(C_1 {+} 3C_2 {+}C_4 \right) \Theta^2
\,. \label{act5}
\end{equation}
This shows explicitly that there are only four independent
Jacobson's coupling constants. Note that the squared acceleration,
shear, vorticity and expansion terms enter the scalar (\ref{act5})
with equal weight (i.e., with equal coefficients), when
$C_2{=}C_3{=}C_4{=}0$ and $C_1$ is free. In this particular
symmetric situation we obtain the case analyzed in \cite{BD09},
namely,
\begin{equation}
K^{abmn}(\nabla_a U_m) (\nabla_b U_n) = C \, \Psi_{ik} \Psi^{ik}
\equiv  C \,\Psi^2 =
C \, \left[DU_k DU^k {+} \sigma_{ik}
\sigma^{ik} {+} \omega_{ik} \omega^{ik} {+} \frac{1}{3} \Theta^2
\right]\,, \label{act6}
\end{equation}
where $C\equiv C_1$.  Thus, we have shown that
the maximal number of
components of the tensor $K^{abmn}$ is 78, Jacobson's theory
admits 4 independent components, and this number
can be reduced to one in the case with high
symmetry mentioned above.

\subsection{Coupling constants related to a spontaneous
polarization-magnetization of the matter or
vacuum induced by an aether
non-uniform motion}
\label{coupb}

\subsubsection{Preliminary analysis}
\label{schemeofcoup}

We now analyze
the spontaneous polarization-magnetization tensor ${\cal H}^{ik}$.
This tensor
splits naturally into four terms (see Eq.~(\ref{E4})). The first,
second, and third terms, have as multipliers the tensors $A^{ikmn}$,
$B^{ikmnpq}$, ${\cal B}^{(ml)sik}$, respectively, which depend on
metric $g_{ik}$, the covariant constant Kronecker tensors
($\delta^i_k$, $\delta^{ik}_{ab}$ and higher order Kronecker tensors),
the Levi-Civita tensor $\epsilon^{ikab}$, and the unit vector field
$U^k$.  The fourth term has $Q$ as a multiplier and due to its
simplicity does not require special consideration. To analyze the
first two terms, i.e., those containing $A^{ikmn}$ and $B^{ikmnpq}$,
we use a decomposition of the polarization ${\cal P}^{m}$ and
magnetization ${\cal M}^{m}$ four-vectors with respect to the
irreducible parts of the covariant derivative of the unit vector field
$U^k$. This approach is useful as it gives a direct method of
interpretation of the corresponding coupling constants. In the
analysis of the third term ${\cal B}^{(ml)sik}$ in Eq.~(\ref{E4}), we
follow another route, as this method of using the decomposition of
${\cal P}^{m}$ and ${\cal M}^{m}$ is not effective since ${\cal
B}^{(ml)sik}$ contains a second covariant derivative of the velocity
four-vector. Instead, we use the standard decomposition of ${\cal
B}^{(ml)sik}$ with respect to the metric $g_{ik}$, the covariant
constant Kronecker tensors ($\delta^i_k$, $\delta^{ik}_{ab}$ and
higher order Kronecker tensors), the Levi-Civita tensor
$\epsilon^{ikab}$, and the unit vector field $U^k$.

Generically, the skew-symmetric tensor ${\cal H}^{ik}$
appearing in Eq.~(\ref{E4})
can be
represented as (see, e.g., \cite{ME})
\begin{equation}
{\cal H}^{ik} =  \delta^{ik}_{mn}U^n {\cal P}^{m} -
\epsilon^{ikmn}U_n {\cal M}_{m}  \,,
\label{P0}
\end{equation}
where ${\cal P}^{m}$ is the spontaneous polarization four-vector and
${\cal M}^{m}$ is the spontaneous magnetization
pseudo four-vector. One can invert the relation (\ref{P0}) and find
\begin{equation}
{\cal P}^{i} \equiv {\cal H}^{ik}U_k  \,,
\quad
{\cal M}^{i} \equiv \frac{1}{2} \epsilon^{ikmn}{\cal H}_{mn}U_k \,.
 \label{M111}
\end{equation}
The four-vectors ${\cal P}^{m}$ and ${\cal
M}_{m}$ are orthogonal to the velocity four-vector, i.e.,
\begin{equation}
{\cal P}^{i} U_i  = 0\,,\quad {\cal M}_{i} U^i = 0   \,,
\label{P3}
\end{equation}
and this fact simplifies the decomposition of these quantities with
respect to irreducible parts of the covariant derivative of the
velocity four-vector (\ref{act3}).
Our scheme of
analysis and interpretation of the coupling constants is the
following: we decompose the four-vectors ${\cal P}^{i}$ and ${\cal
M}^{i}$ with respect to $DU^i$, $\sigma^{ik}$, $\omega^{ik}$, $\Theta
\Delta^{ik}$ using unknown coupling constants, and then reconstruct
${\cal H}^{ik}$ using (\ref{P0}).

Note that the term spontaneous, in spontaneous polarization and
spontaneous magnetization, is being used following the terminology of
classical electrodynamics. One usually distinguishes between
non-spontaneous polarization or non-spontaneous magnetization induced
by an electromagnetic field on one hand, and the spontaneous
polarization or spontaneous magnetization of non-electromagnetic
origin, produced, e.g., by medium deformation or heating, on the other
hand.  In this sense, the term spontaneous is appropriate for
spontaneous polarization or spontaneous magnetization caused by the
interaction of the medium with a non-uniformly moving aether.

\subsubsection{Reconstruction of the tensors $A^{mnik}$ and
$B^{mnlsik}$ } \label{reconstructionAB}

\vskip 0.3cm
\noindent
{\it (a) Polarization-magnetization linear in the covariant
derivative of the unit vector field, ${\cal
H}^{ik}_{(1)} {=} A^{mnik} \nabla_m U_n$}

\noindent
Let us start with the analysis of the first term
in Eq.~(\ref{E4}),
\begin{equation}
{\cal
H}^{ik}_{(1)} {=} A^{mnik} \nabla_m U_n  \,,
\label{hij0}
\end{equation}
that contributes to the tensor of spontaneous
polarization-magnetization.  Studying the symmetry of the tensor
coefficients $A^{ikmn}$, appearing in Eq.~(\ref{E4}), and in analogy
with $K^{abmn}$, we find that $A^{ikmn} {=} {-} A^{iknm}$, since the
Maxwell tensor is skew-symmetric. Again, we can also put $A^{ikmn}U_m
{=}0$, since $U^m \nabla_i U_m{=}0$.  Formally speaking, there are, in
general, $6 \times 12 {=} 72$ independent components of the tensor
$A^{ikmn}$. Nevertheless, we intend to show that using the unit vector
field $U^m$ itself and the geometric quantities $g^{ik}$,
$\delta^i_k$, $\delta^{ik}_{pq}$, $\epsilon^{ikmn}$, we can
reconstruct this tensor based on two and only two independent coupling
constants.

Searching for first order, i.e., linear, terms in the decomposition of
${\cal P}^{i}$ we can find only one natural four-vector, $DU^i$, and
for the decomposition of ${\cal M}^{i}$ we can find only one natural
pseudo four-vector, $\omega^{*i}{=} \omega^{*ik}U_k$. Thus, the
corresponding first order decompositions are
\begin{equation}
{\cal P}^{i}_{(1)} =  \pi_1 DU^i  \,, \quad {\cal M}^{i}_{(1)} =
\mu_1 \epsilon^{ikpq} U_k \omega_{pq} \,,
\label{P000}
\end{equation}
with $\pi_1$ and $\mu_1$ being independent coupling constants.

\vskip 0.3cm

\noindent
{\it (b)
Polarization-magnetization quadratic in the covariant
derivative of the unit vector field,
${\cal H}^{ik}_{(2)} =  B^{mnlsik} (\nabla_l U_s)(\nabla_m U_n)$}

\noindent
Let us now analyze the second term in Eq.~(\ref{E4}),
\begin{equation}
{\cal H}^{ik}_{(2)} =  B^{mnlsik} (\nabla_l U_s)(\nabla_m U_n)\,.
\label{secter}
\end{equation}
The tensor $B^{ikmnpq}$ possesses the following symmetries
\begin{equation}
B^{mnlsik} = - B^{mnlski} = B^{lsmnik} \,, \quad B^{mnlsik}U_n =
0\,,\quad  B^{mnlsik}U_s=0 \,. \label{M60}
\end{equation}
Thus, in general, $B^{ikmnpq}$
can be characterized by $\frac12 \times 12 \times
13 \times 6 {=}468$ independent components.
Nevertheless, below we
show that the reconstruction of the tensor $B^{ikmnpq}$ requires
the introduction of only five coupling constants.

To deal then with second-order terms, represented by ${\cal
H}^{ik}_{(2)} =  B^{mnlsik} (\nabla_l U_s)(\nabla_m U_n)$, we find
that there are the following quadratic terms: $\Theta^2$, $ DU^i\,
DU^j$, $\sigma_{ik} \, \sigma_{mn}$, $ \omega_{ik} \,\omega_{mn}$.
With these quantities we cannot construct neither a vector, nor a
pseudo vector orthogonal to $U^i$. There are also quadratic
cross-terms, namely, $DU^i \, \Theta$, $DU^i \, \sigma_{mn}$,
$DU^i \, \omega_{mn}$, $\Theta \, \omega_{ij}$, $\sigma_{ij}
\,\omega_{mn}$ and $\Theta \,\sigma_{ij}$. The first term, $DU^i
\, \Theta$, is a four-vector, the next two terms, i.e., $DU^i \,
\sigma_{mn}$, and $DU^i \, \omega_{mn}$, can be contracted to form
vectors that enter into the decomposition of ${\cal P}^{i}$. The
next two terms, i.e., $\Theta \, \omega_{ij}$ and $\sigma_{ij}
\,\omega_{mn}$, can be contracted with the Levi-Civita tensor and
with the velocity four-vector to form pseudo vectors that enter
into the decomposition of ${\cal M}^{i}$. The last term, i.e.,
$\Theta \,\sigma_{ij}$, cannot form a vector or pseudo vector in
any way, as the trace of the symmetric tensor $\sigma_{ij}$ is
zero.

Thus, in summary,
there are only one linear and three  quadratic
terms in the decomposition of the polarization
four-vector, ${\cal P}^{i}$. Therefore
${\cal P}^{i}$ can be written as
\begin{equation}
{\cal P}^{i} = DU_k \left[g^{ik} \left(\pi_1 + \pi_2 \Theta
\right) + \pi_3 \sigma^{ik} + \pi_4 \omega^{ik} \right] \,,
\label{P4}
\end{equation}
with $\pi_1$, $\pi_2$, $\pi_3$, and $\pi_4$ being four independent
coupling constants.  An interesting
aspect of this representation is
that the polarization four-vector ${\cal P}^{i}$ is
proportional to the acceleration four-vector
$DU_k$, and so ${\cal P}^{i}$
vanishes when $DU_k$ is
equal to zero, $DU_k{=}0$.

Similarly, we can construct ${\cal M}^{i}$ with one linear term and
three quadratic terms. ${\cal M}^{i}$ can then be written as
\begin{equation}
{\cal M}^{i} {=} \epsilon^{ikpq} U_k \omega_{nm}\left[
\delta^n_p \delta^m_q \left(\mu_1 {+} \mu_2 \Theta \right)
{+}  \mu_3 (\sigma^{m}_p \delta_q^n{-}\sigma^{m}_q
\delta^n_{p})\right]
\,,
\label{M4}
\end{equation}
where $\mu_1$, $\mu_2$, and $\mu_3$, are 3 new independent coupling
constants.  There is no magnetization when there is no vorticity,
i.e., $\omega_{nm}{=}0$.  One can also indeed add another term to the
right hand side of Eq.~(\ref{M4}), namely, $\mu_4 \Delta^i_l \
\epsilon^{lkpq} \ \omega_{pq} DU_k$, in which case ${\cal M}^{i}$ gets
another coupling constant $\mu_4$.  However, when we reconstruct the
tensor ${\cal H}^{ik}$ in (\ref{P0}), the term with $\mu_4$ disappears
because of the properties of products of two Levi-Civita symbols
together with the contractions of $U^s$ with $\omega_{pq}$ and $DU_k$.
Thus $\mu_4$ is a hidden coupling constant which does not enter into
the dynamics, so it can be put to zero without loss of generality,
$\mu_4 {=}0$.

\vskip 0.3cm
\noindent
{\it (c)
{Reconstruction of the tensors $A^{mnik}$ and $B^{mnlsik}$}}

\noindent
Now we put Eqs.~(\ref{P4}) and (\ref{M4}) into
Eq.~(\ref{P0}) and compare the
result with Eq.~(\ref{E4}). It is then
possible to reconstruct the tensor
$A^{mnik}$, namely,
\begin{equation}
A^{mnik} = \pi_1 g^{iknl} U^m U_l - \mu_1 \Delta^{ikmn} \,,
\label{M5}
\end{equation}
where we have introduced the following auxiliary tensors
\begin{equation}
g^{mnpq} \equiv g^{mp} g^{nq}{-}g^{mq} g^{np} \,,
\label{M32}
\end{equation}
\begin{equation}
\Delta^{mnpq} \equiv \Delta^{mp} \Delta^{nq}{-}\Delta^{mq}
\Delta^{np} \,.
\label{M321}
\end{equation}
Similarly, we can reconstruct the tensor $B^{mnlsik}$.
It is given by
$$
B^{mnlsik} = \frac12 \pi_2 U_p \left[\Delta^{mn}
U^l g^{iksp} {+} \Delta^{ls} U^m g^{iknp} \right] {+}
$$
$$
{+}\frac12 \pi_3 \delta^{[i}_p U^{k]}\left\{U^l \left[\Delta^{m(p}
\Delta^{s)n} {-} \frac13 \Delta^{ps}\Delta^{mn} \right] {+} U^m
\left[\Delta^{l(p} \Delta^{n)s} {-} \frac13 \Delta^{pn}\Delta^{ls}
\right]\right\} +
$$
$$
+ \frac14 \pi_4 \delta^{[i}_p U^{k]}\left(U^l \Delta^{psmn} + U^m
\Delta^{pnls} \right) - \frac12 \mu_2 \left(\Delta^{mn}
\Delta^{ikls}+ \Delta^{ls} \Delta^{ikmn}\right) -
$$
\begin{equation}
- \frac12 \mu_3 \left[ \frac23 \left(\Delta^{mn} \Delta^{ikls} {+}
\Delta^{ls}\Delta^{ikmn}\right) {+} \Delta^{n[i} \Delta^{k]mls}
{+}  \Delta^{m[i} \Delta^{k]nls} {+} \Delta^{s[i} \Delta^{k]lmn} +
\Delta^{l[i} \Delta^{k]smn}\right] \,. \label{M96}
\end{equation}
The tensor
$A^{mnik}$ contains two coupling constants,
and the tensor $B^{mnlsik}$
contains five coupling constants.

\subsubsection{Polarization-magnetization associated with the second
covariant derivative of the unit vector field}
\label{ploarmagsecondder}

Now we analyze the third term in Eq.~(\ref{E4}), namely,
\begin{equation}
{\cal H}^{ik}_{({\rm DD})} \equiv  {\cal B}^{(ml)sik}
\nabla_{(m} \nabla_{l)} U_s\,.
\label{PMnon3}
\end{equation}
The tensor  ${\cal B}^{(ml)sik}$ is symmetric with respect to
the indices $(ml)$. Generally it
possesses $\frac12 \times 4 \times 5
\times 4 \times 6 {=} 240$ independent components, and thus can be
described using 240 independent coupling constants.
However, in fact five
coupling constants are enough to represent this tensor using
the
metric $g_{ik}$, the Kronecker tensors
($\delta^i_k$, $\delta^{ik}_{ab}$ and higher order Kronecker tensors),
the Levi-Civita tensor $\epsilon^{ikab}$, and the unit vector field
$U^k$.

The tensor  ${\cal B}^{(ml)sik}$
cannot be decomposed as was done for the
terms ${\cal H}^{ik}_{(1)}$ and
${\cal H}^{ik}_{(2)}$. We use here another approach.
Keeping in mind
the symmetry of the tensor ${\cal B}^{(ml)s[ik]}$ and  that
there are only two natural symmetric tensors,  i.e.,
$g^{ik}$ and $U^i U^k$,
and only two natural skew-symmetric pure tensors, i.e.,
$\delta^{ik}_{pq}$
and $\Delta^{ikmn}$, this tensor has to be of the form
\begin{equation}
{\cal B}^{(ml)sik} = \delta^{ik}_{pq} U^q \left[g^{ps}\left(\rho_1
g^{ml} {+} \rho_2 U^{m} U^{l} \right) {+} \rho_3 g^{p(m}g^{l)s}
{+} \rho_4 U^s g^{p(m} U^{l)} \right] {+} \rho_5
\Delta^{iks(m}U^{l)} \,, \label{PMnon4}
\end{equation}
where five new
coupling constants $\rho_1$, $\rho_2$,
$\rho_3$, $\rho_4$, and $\rho_5$
have appeared.

\subsubsection{Nonminimal polarization-magnetization}
\label{nonminimalpolarmag}

The nonminimal part of the polarization-magnetization tensor
\begin{equation}
{\cal H}^{ik}_{({\rm NM1})} \equiv  Q U_m R^{m[i}U^{k]}
\label{PMnon1}
\end{equation}
is associated with the vanishing magnetization four-vector
${\cal M}^i_{({\rm NM})}{=}0$ and the polarization four-vector of the form
\begin{equation}
{\cal P}^{i}_{({\rm NM})} = \frac12 Q U_m R^{mk} \Delta^i_k \,.
\label{PMnon2}
\end{equation}
Only one coupling constant, $Q$, describes the polarization of the
medium/vacuum, induced by the interaction with curvature in the presence
of unit vector field.

\subsection{Coupling constants related to
the permittivity tensors of the matter or vacuum
in a non-uniformly moving aether}
\label{xandy}

\subsubsection{Susceptibilities linear and quadratic in the covariant
derivatives of the unit vector field}
\label{susceplinequadrat}

In order to represent the tensors $X^{lsikmn}$ and $Y^{ablsikmn}$ in
$C_{({\rm D})}^{ikmn}$ (see (\ref{E7})), we use a
similar scheme, as
for the coefficients $A^{mnik}$ and $B^{mnlsik}$.
We start with the linear response tensor $C^{ikmn}$ given in
Eq.~(\ref{E5}). It admits the standard decomposition
$$
C^{ikmn} =  \frac12 \left( \varepsilon^{im} U^k U^n {-}
\varepsilon^{in} U^k U^m {+} \varepsilon^{kn} U^i U^m {-}
\varepsilon^{km} U^i U^n \right)  - \frac12
\eta^{ikl}(\mu^{-1})_{ls}  \eta^{mns} +
$$
\begin{equation}
+\frac12 \left[\eta^{ikl}(U^m\nu_{l}^{\
n} {-} U^n \nu_{l}^{\ m}) {+} \eta^{lmn}(U^i \nu_{l}^{\ k} {-} U^k
\nu_{l}^{\ i} ) \right] \,, \label{44}
\end{equation}
where $\varepsilon^{im}$ is the dielectric permittivity tensor,
$(\mu^{-1})_{pq}$ is the magnetic impermeability tensor,
$\nu_{p \ \cdot}^{\ m}$ is the tensor of magneto-electric
coefficients, i.e.,
\begin{equation}
\varepsilon^{im} = 2 C^{ikmn} U_k U_n \,, \quad (\mu^{-1})_{pq}  =
- \frac{1}{2} \eta_{pik}  C^{ikmn} \eta_{mnq}\,, \quad \nu_{p}^{\
m} = \eta_{pik} C^{ikmn} U_n =U_k C^{mkln} \eta_{lnp}\,.
\label{varco}
\end{equation}
As usual, the tensors $\eta_{mnl}$ and $\eta^{ikl}$ are
skew-symmetric tensors orthogonal to $U^i$,
\begin{equation}
\eta_{mnl} \equiv \epsilon_{mnls} U^s \,,
\quad
\eta^{ikl} \equiv \epsilon^{ikls} U_s \,,
\label{47}
\end{equation}
and obey the following identities
\begin{equation}
- \eta^{ikp} \eta_{mnp} = \delta^{ikl}_{mns} U_l U^s = \Delta^i_m
\Delta^k_n - \Delta^i_n \Delta^k_m \,, \quad -\frac{1}{2}
\eta^{ikl}  \eta_{klm} = \delta^{il}_{ms} U_l U^s =  \Delta^i_m
\,. \label{501}
\end{equation}
We now decompose
explicitly the permittivity tensors $\varepsilon^{im}$,
$(\mu^{-1})_{pq}$ and $\nu^{pm}$ using the irreducible parts of the
covariant derivative of the velocity four-vector
(namely, $DU^i$,
$\sigma_{ik}$, $\omega_{pq}$, and $\Theta$).
The properties
\begin{equation}
\varepsilon_{ik}U^k = 0 \,, \quad {(\mu^{{-}1})}^{ik} U_k = 0 \,,
\quad \nu^{ik}U_k = 0 = \nu^{ik}U_i \label{ortho}
\end{equation}
simplify the decomposition of $\varepsilon^{im}$,
$(\mu^{-1})_{pq}$ and $\nu^{pm}$, and the results are the following.
The dielectric permittivity tensor is decomposed as
$$
\varepsilon^{ik} = \Delta^{ik} \left(\varepsilon {+} \alpha_1
\Theta \right){+} \Delta^{ik} \left(\alpha_2 DU_m DU^m {+}
\alpha_3 \Theta^2 {+} \alpha_4 \sigma_{mn}\sigma^{mn} {+} \alpha_5
\omega_{mn}\omega^{mn} \right) {+}
$$
\begin{equation}
{+} \alpha_6 \sigma^{ik} {+} \alpha_7 \Theta \sigma^{ik} {+}
\alpha_8 DU^i DU^k {+} \alpha_9 \sigma^{ip} \sigma^k_p {+}
\alpha_{10} \ \omega^{ip}\omega^k_{\ p} {+} \alpha_{11} \
\sigma^{(i}_{p}\omega^{k)p} \,, \label{eps}
\end{equation}
where  $\varepsilon$ and $\alpha_1, ...,\alpha_{11}$
form
twelve independent coupling constants.
The magnetic impermeability tensor is decomposed as
$$
{\left(\mu^{-1}\right)}^{ik} = \Delta^{ik} \left(\frac{1}{\mu}{+}
\gamma_1 \Theta \right) {+} \Delta^{ik} \left[\gamma_2 DU_m DU^m
{+} \gamma_3 \Theta^2 {+} \gamma_4 \sigma_{mn}\sigma^{mn} {+}
\gamma_5 \omega_{mn}\omega^{mn} \right] +
$$
\begin{equation}
{+} \gamma_6 \sigma^{ik} {+} \gamma_7 \Theta \sigma^{ik} {+}
\gamma_8 DU^i DU^k {+} \gamma_9 \sigma^{ip} \sigma^k_p {+}
\gamma_{10} \omega^{ip}\omega^k_{\ p} {+} \gamma_{11}
\sigma^{(i}_{p}\omega^{k)p}\,, \label{mu}
\end{equation}
where  $\mu$ and $\gamma_1, ...,\gamma_{11}$ form also
twelve independent coupling constants.
The magneto-electric cross-effect pseudo tensor is decomposed as
$$
\nu^{pm} = (\nu_1{+} \nu_3 \Theta) \Delta^p_q
\Delta^m_n \omega^{*qn} {+}(\nu_2 {+} \nu_4 \Theta)
\eta^{pml} DU_l {+}
$$
$$
{+}\nu_5 \Delta^{s(p} \omega^{*}_{sq} \sigma^{m)q} {+}\nu_6
\Delta^{s[p} \omega^{*}_{sq} \sigma^{m]q}{+} \nu_7 \
\omega^{(p}_{\ \ q} \Delta^{m)}_s \omega^{*sq} {+} \nu_8 \
\omega^{[p}_{\ \ q} \Delta^{m]}_s \omega^{*sq}{+}
$$
\begin{equation}
{+}\nu_9 DU^{(p} \omega^{*m)q} U_q  {+}\nu_{10} DU^{[p}
\omega^{*m]q} U_q {+}\nu_{11} \sigma^{q(p} \eta^{m)}_{\ \ \ ql}
DU^l {+}\nu_{12} \sigma^{q[p} \eta^{m]}_{\ \ \ ql} DU^l \,,
\label{nu}
\end{equation}
where  $\nu_1, ...,\nu_{12}$ form
another twelve independent coupling constants.

Having decomposed explicitly the permittivity tensors
$\varepsilon^{im}$, $(\mu^{-1})_{pq}$ and $\nu^{pm}$ using the
irreducible parts of the covariant derivative of the velocity
four-vector we can now reconstruct the tensors $X^{lsikmn}$ and
$Y^{ablsikmn}$ in $C_{({\rm D})}^{ikmn}$ given in Eq.~(\ref{E7}),
keeping in mind their symmetry,
\begin{equation}
X^{lsikmn}=-X^{lskimn}=-X^{lsiknm}=X^{lsmnik} \,, \label{49}
\end{equation}
\begin{equation}
Y^{ablsikmn}=-Y^{ablskimn}=Y^{ablsmnik}=Y^{lsabikmn} \,.
\label{50}
\end{equation} The reconstructed tensors $X^{lsikmn}$ and
$Y^{ablsikmn}$ are presented in Appendix B.

The given representation of the permittivity tensors allows us to
interpret and classify the coupling constants appearing
in this decomposition.
Two constants, $\varepsilon$ and
$\mu$, have a standard interpretation in terms of an aether uniform
motion. Other coupling constants can be classified
with respect to electrodynamic effects which can exist when
the aether is in a state
of non-uniform motion.
For instance, the magneto-electric coefficients, described by
the non-symmetric tensor $\nu^{pm}$, represent the effect of optical
activity, and it can be splitted into a sum of symmetric and
skew-symmetric parts. Thus, the term with the coupling constant
$\nu_1$ is related to the polarization rotation phenomenon linear in
the vorticity tensor $\omega^{pm}$ and is purely skew-symmetric
contribution to the linear term. Similarly, the parameters $\nu_7$ and
$\nu_8$ relate to quadratic effects, symmetric and skew-symmetric,
respectively.  The coupling constant $\nu_2$ is connected with the
optical activity caused by an acceleration of the aether with the
effect being linear in $DU^i$. The parameters $\nu_3, ..., \nu_6$ and\
$\nu_9,..., \nu_{12}$ are connected to the corresponding
cross-effects.

\subsubsection{Susceptibilities containing second covariant derivatives
of the unit vector field}
\label{suscepunitvec}

The term $C_{({\rm DD})}^{ikmn} = {\cal Y}^{jlsikmn} \nabla_{(j}
\nabla_{l)} U_s$ in Eq.~(\ref{E73}) relates to a linear response
induced by a second covariant derivative of the unit vector field
$U^i$.  The tensor ${\cal Y}^{jlsikmn}={\cal Y}^{(jl)s[ik][mn]}$, in
general, possesses $\frac12 \times 4 \times 5 \times 4 \times \frac12
\times 6 \times 7 {=} 840$ components. In our setting it can be
characterized by twelve coupling constants only, and similarly to the
tensor ${\cal B}^{(ml)s[ik]}$ in Eq.~(\ref{PMnon4}), it can be
represented as follows,
$$
{\cal Y}^{jlsikmn} {=} g^{ikmn} \left[U^s \left(\rho_{6} g^{jl}{+}
\rho_{7} U^j U^l \right){+} \rho_{8} g^{s(j}U^{l)}  \right]{+}
\rho_{9} U^s \left(g^{ikj[n} g^{m]l} {+} g^{ikl[n} g^{m]j}\right)
{+}
$$
$$
{+}\rho_{10} \left[U^l \left(g^{iks[m} g^{n]j}{+}g^{mns[i}
g^{k]j}\right) {+}U^j \left(g^{iks[m} g^{n]l} {+} g^{mns[i}
g^{k]l}\right)\right]
$$
$$
{+} \rho_{11} U^s \left(g^{ikj[m} U^{n]} U^l {+}g^{ikl[m} U^{n]}
U^j {+}g^{mnj[i} U^{k]} U^l {+}g^{mnl[i} U^{k]} U^j \right) {+}
$$
$$
{+} \rho_{12} U^j U^l \left(g^{iks[m} U^{n]} {+} g^{mns[i}
U^{k]}\right){+} \left(\rho_{13} g^{jl}U^s {+}\rho_{14}g^{s(j}
U^{l)} {+} \rho_{15} U^s U^j U^l\right)U^{[i} g^{k][m} U^{n]}{+}
$$
\begin{equation}
{+} \rho_{16}U^s \left(g^{j[i}U^{k]} g^{l[m} U^{n]}{+}
g^{l[i}U^{k]} g^{j[m} U^{n]}\right){+} \rho_{17} \left(U^{(l}
g^{j)[i} U^{k]} g^{s[m} U^{n]} {+} U^{(l} g^{j)[m} U^{n]} g^{s[i}
U^{k]}\right)\,. \label{DD0}
\end{equation}

\subsubsection{Nonminimal susceptibilities}
\label{suscepnonmin}

This term is given in Eq.~(\ref{E73}) as $C_{({\rm R})}^{ikmn} {=}
{\cal R}^{ikmn} + S^{pqabikmn} R_{pqab}$, and is a term that
contributes to the total linear response tensor (\ref{E5}).
According to Eq.~(\ref{sus1}) the nonminimal susceptibility tensor
${\cal R}^{ikmn}$ contains three independent coupling constants,
namely, $q_1$, $q_2$ and $q_3$. The tensor $S^{pqabikmn}$ is
skew-symmetric with respect to indices $pq$, $ab$, $ik$, $mn$,  is
symmetric with respect to transpositions $pq \to ab$, $ik \to mn$,
and thus can be characterized, in general, by $\frac12 \times 6
\times 7 \times \frac12 \times 6 \times 7 {=} 441$ components.
When we only use the metric $g_{ik}$, the Kronecker tensors
($\delta^i_k$, $\delta^{ik}_{ab}$ and higher order Kronecker
tensors), the Levi-Civita tensor $\epsilon^{ikab}$, and the unit
vector field $U^k$, in its reconstruction, this tensor has 11
independent terms, so 11 coupling constants. It has thus the form,
see Appendix~\ref{App:AppendixB}  for details,
\begin{equation}
S^{pqabikmn} {=}S^{pqabikmn}_{(1)}{+}
S^{pqabikmn}_{(2)}{+}S^{pqabikmn}_{(3)} \,,
\label{S000}
\end{equation}
where
\begin{equation}
S^{pqabikmn}_{(1)} {=} \left(q_4 g^{pqab} {+} q_5
\Delta^{pqab}\right) \left(\Delta^{ikmn}{-}g^{ikmn}\right) {+}
q_6 g^{ikmn}\left(U^{[p}g^{q][a} U^{b]} {+} U^{[a}g^{b][p}
U^{q]}\right) \,, \label{S001}
\end{equation}
$$
S^{pqabikmn}_{(2)} {=} q_7 \left(U^{[a}
g^{b][p}g^{q][m}g^{n][i}U^{k]} {+} U^{[a}
g^{b][p}g^{q][i}g^{k][m}U^{n]} {+}U^{[p}
g^{q][a}g^{b][m}g^{n][i}U^{k]} {+} U^{[p}
g^{q][a}g^{b][i}g^{k][m}U^{n]} \right) {+}
$$
$$
{+}q_8 \left(U^{[p} g^{q][i}g^{k][m}g^{n][a}U^{b]} {+} U^{[p}
g^{q][m}g^{n][i} g^{k][a}U^{b]} {+}U^{[a}
g^{b][i}g^{k][m}g^{n][p}U^{q]} {+} U^{[a}
g^{b][m}g^{n][i}g^{k][p}U^{q]} \right) {+}
$$
$$
{+}q_9 \left(U^{[p} g^{q][i} U^{k]} U^{[m} g^{n][a} U^{b]} {+}
U^{[p} g^{q][m} U^{n]} U^{[i} g^{k][a} U^{b]} {+} U^{[a} g^{b][i}
U^{k]} U^{[m} g^{n][p} U^{q]} {+} U^{[a} g^{b][m} U^{n]} U^{[i}
g^{k][p} U^{q]} \right){+}
$$
\begin{equation}
{+} q_{10} \left(g^{abmn} U^{[i}g^{k][p} U^{q]} {+} g^{abik}
U^{[m}g^{n][p} U^{q]} {+} g^{pqmn} U^{[i}g^{k][a} U^{b]} {+}
g^{pqik} U^{[m}g^{n][a} U^{b]} \right) \,, \label{S00}
\end{equation}
and
the last term $S^{pqabikmn}_{(3)}$ can
be obtained from $S^{pqabikmn}_{(2)}$
with introduction of new coupling constants $q_{11}, ..., q_{14}$
instead of $q_{7}, ..., q_{10}$, respectively,
and by the substitution
\begin{equation}
U^{[i}g^{k]p} \equiv - g^{ikpq} U_q \to \epsilon^{ikpq}U_q \,.
\label{S012}
\end{equation}
Clearly, the tensor $S^{pqabikmn}$ possesses the required symmetries,
and
the unit vector field enters this quantity in even combinations of
second and fourth orders.

\subsection{Summary of the decompositions}
\label{sumup}

Let us sum up the independent parameters of
the Einstein-Maxwell-aether theory
we are interested in. We started with the four parameters
$C_1,C_2.C_3,C_4$ introduced in the pure Einstein-aether theory.  Then we
added thirteen parameters $\pi_1,\pi_2, \pi_3,\pi_4$,
$\mu_1,\mu_2,\mu_3$,
$Q$, $\rho_1,\rho_2,\rho_3,\rho_4,\rho_5$
appearing in the decomposition of the
spontaneous polarization-magnetization tensor.  Also there are
thirty-six coupling constants $\varepsilon, \mu$, $\alpha_1, ...,
\alpha_{11}$, $\gamma_1, ..., \gamma_{11}$, and $\nu_1, ...,
\nu_{12}$, and
fourteen nonminimal coupling constant
parameters $q_1,...,q_{14}$. Finally, we have introduced twelve
parameters $\rho_6,...,\rho_{17}$. In total the theory has a set of 79
independent parameters.

\subsection{Three spacetime models with high symmetry:
Remarks on the structure of the unit vector field
$U^i$ based on the analysis of
the compatibility conditions}
\label{remunitvector}

\subsubsection{Motivation}

Keeping in mind applications of this Einstein-Maxwell-aether theory,
we would like to call the attention to three interesting consequences
coming from the analysis of the structure of the unit vector field.
Indeed, three spacetime models with high symmetry are prone to be
solutions, possibly analytical solutions, of the
Einstein-Maxwell-aether theory presented here. These spacetimes models
are the spatially homogeneous cosmological models,
 static spherically symmetric structures, and plane-wave spacetimes.
We do not intend here to
analyze the total system of reduced master equations, but would like
to mention the consequences, which follow from the compatibility
conditions related to our ansatz on the structure of the unit vector
field.

\subsubsection{Three spacetime models}

\noindent {\it (a) Spatially homogeneous cosmological models}
\label{spatialhomocosmolog}

\noindent Let us consider first, the
Friedmann-Lema\^itre-Robertson-Walker (FLRW) cosmological models
with line element
\begin{equation}
ds^2 = dt^2 - a^2(t)\left( dx^2 + dy^2 + dz^2\right) \,,
\label{S1}
\end{equation}
where $a(t)$ is the Friedmann scale factor as a function
of the cosmological time $t$, and
$x,y,z$ are spatial homogeneous coordinates.
Within these models we can assume that the aether velocity
four-vector is of the form $U^i {=} \delta^i_t$, and thus the tensor
$\nabla_m U_n$ has the following irreducible terms
\begin{equation}
U_m\,DU_n = 0, \ \sigma_{ik}= 0,  \ \omega_{pq} = 0, \ \Theta =
3\,\frac{\dot{a}}{a} = 3H(t) \,, \label{S2}
\end{equation}
where $H(t)\equiv\frac{\dot{a}}{a}$ is the Hubble function.
In such a case we find, that ${\cal P}^i {=} 0\,$, ${\cal M}^i {=} 0$ and
\begin{equation}
K^{abmn}(\nabla_aU_m) (\nabla_b U_n) = \frac13
\Theta^2 (C_1{+} 3C_2 {+} C_3) \,.
\label{S3}
\end{equation}
The spacetime symmetries require that the global electromagnetic field
obeys $F_{ik}{=}0$, and the corresponding electrodynamic equations
are satisfied identically, since $I^i{=}0$, ${\cal P}^i = 0$ and
${\cal M}^i = 0$.  We obtain the standard FLRW cosmological model, if
we prove that the equations for the aether velocity are satisfied
identically, when $U^i {=}\delta^i_t$.  Indeed, ${\cal J}^{mn}_{({\rm
M})}{=}0$ and $I^n_{({\rm M})}{=}0$, since $F_{pq}{=}0$. If we suppose
that $\frac{\delta L_{({\rm m})}}{\delta U_n}{=}0$, then, $I^n_{({\rm
A})}{=}0$, since $DU^n{=}0$.  The term ${\cal J}^{mn}_{({\rm A})}$
yields
\begin{equation}
{\cal J}^{mn}_{({\rm A})} = \frac13 \Theta
\left[(C_1{+}3C_2{+}C_3)g^{mn} - (C_1{+}C_3)U^m U^n \right]\,,
\label{S5}
\end{equation}
and the reduced equation (\ref{A2})
\begin{equation}
\Delta_n^s \nabla_m {\cal J}^{mn}_{({\rm A})} = 0 \,,
\label{S6}
\end{equation}
is satisfied identically. Thus we have
checked that in the spatially
homogeneous FLRW cosmological models without a global
electromagnetic field, the aether coupling parameters remain
hidden, the unit vector field being of the form $U^i {=}\delta^i_t$.

A non-uniform aether motion may provide the appearance of
unlighted cosmological epochs similar to the ones described in
\cite{BBL12}. In these
unlighted epochs, the square of the effective refraction
index is negative, and the corresponding electromagnetic waves can not
propagate.

\vskip 0.5cm \noindent
{\it (b) Static spherically symmetric
models} \label{staticspherical}

\noindent
We now assume a static spherically symmetric metric spacetime
with line element
\begin{equation}
ds^2 = B(r)dt^2 - A(r)dr^2 - r^2(d\theta^2
+ \sin^2{\theta} d\varphi^2) \,,
\label{SS1}
\end{equation}
where $t$ is the global time, $(r,\theta,\phi)$
are the spherical symmetric spatial coordinates, and
$B(r)$, $A(r)$ are the metric functions.
Let us assume that the aether velocity four-vector is of the form
\begin{equation}
U^i{=}\delta^i_t
\frac{1}{\sqrt{B(r)}}\,.
\label{SSSaether}
\end{equation}
This assumption, that the
aether is aligned with the timelike Killing vector is not
the most general,
and can be put under scrutiny on physical
grounds, as in general the aether has radial and time
components as it falls into a central body, see
\cite{J42,J7,bbm} for a more general
class of spherical symmetric solutions.
Nevertheless, we maintain here the
assumption given in Eq,~(\ref{SSSaether})
and leave
for another work the study of
more general examples of exact spherically symmetric solutions
to the Einstein-Maxwell-aether theory.

The irreducible parts of the
covariant derivative are then
\begin{equation}
U_mDU_n = {-} \frac{B^{\prime}}{2B} \delta^r_n\, U_m\,, \quad
\sigma_{ik} = 0 \,, \quad \omega_{pq} = 0 \,, \quad \Theta = 0 \,,
\label{SS2}
\end{equation}
where a ${}^{\prime}$ means a derivative with respect to $r$.
The reduced quantity $K^{abmn}(\nabla_aU_m)(\nabla_b U_n)$ is given by
\begin{equation}
K^{abmn}(\nabla_aU_m) (\nabla_b U_n) = (C_1+C_4)DU_m DU^m \,.
\label{SS3}
\end{equation}
In this case there is no magnetization, ${\cal M}^i{=}0$. The
polarization ${\cal P}^i$ four-vector is non-vanishing, its linear
part being of the form ${\cal P}^i {=} \pi_1 DU^i$, and thus
contains the radial component ${\cal P}^r$ only.  The
compatibility conditions for the electrodynamic equations require
then that a static radial electric field should appear in the
system, $E_{\rm radial} \equiv \sqrt{AB} \ F^{r0} \neq 0$, which
in turn is supported by the polarization induced by the aether
non-uniform state. Concerning the gravitational field equations,
one sees that they can be reduced to a pair of equations for
$A(r)$ and $B(r)$, but here we do not intend to specify this set
of equations.

The compatibility of the model as a whole depends on the question of
whether the equation for the aether velocity four-vector $U_t {=}
\sqrt{B}$ is satisfied identically. In fact, in this case one obtains
\begin{equation}
{\cal J}^{mn}_{({\rm A})} =  \left[(C_1+C_4)\delta^m_t \delta^n_r
+C_3 \delta^m_r \delta^n_t   \right] \frac{B^{\prime}}{2AB
\sqrt{B}}\,, \quad \Delta_n^s \nabla_m {\cal J}^{mn}_{({\rm A})} =
0 \,, \quad I^n_{({\rm A})} =  \kappa \frac{\delta L_{({\rm
m})}}{\delta U_n} = 0 \,. \label{SS9}
\end{equation}
Only the equation for $n{=}r$
\begin{equation}
\nabla_m {\cal J}^{mr}_{({\rm M})} = I^r_{({\rm M})} \,,
\label{SS10}
\end{equation}
needs to be analyzed.  Eq.~(\ref{SS10}) can be
reduced to an
identity when ${\cal J}^{rr}_{({\rm M})}{=}0$ and
$I^r_{({\rm M})}{=}0$. This is possible, e.g., for a special
choice of the coupling parameters. We will return to this
problem in the future.

\vskip 0.5cm \noindent {\it (c) Spacetimes with plane-wave
symmetry} \label{planewave}

\noindent
As an illustration for this class of spacetimes we
can consider the metric
\begin{equation}
ds^2 = 2 du dv - L^2 \left(e^{2\beta} {dx^2}^2 +
e^{-2\beta} {dx^3}^2
\right) \,,
\label{PW1}
\end{equation}
where $u$ and $v$
are the retarded and advanced times, respectively,
given in terms of the time $t$ and spatial coordinate $x^1$
by $u{=}\frac{1}{\sqrt2}(t{-}x^1)$, $v{=}\frac{1}{\sqrt2}(t{+}x^1)$,
and $x^2, x^3$ are the other spatial coordinates.
$L(u)$ and
$\beta(u)$ are  functions of the retarded time $u$ only. When the
aether velocity four-vector is assumed to be of the form
\begin{equation}
U^i = \frac{1}{\sqrt2} \left(\delta^i_u +
\delta^i_v \right) = \delta^i_t \,,
\label{PW2}
\end{equation}
i.e., the aether is at rest in the
spacetime reference frame,   the covariant
derivative of the velocity four-vector reduces to
the following equation
\begin{equation}
\nabla_i U^k  = \frac{1}{\sqrt2} \left[\delta_i^2
\delta^k_2 \left(\frac{L^{\prime}}{L}{+}
\beta^{\prime} \right) {+} \delta_i^3 \delta^k_3
\left(\frac{L^{\prime}}{L}{-} \beta^{\prime} \right)\right]\,,
\label{GW001}
\end{equation}
where a prime here denotes a derivative with respect to
the retarded time $u$.
Thus we obtain
\begin{equation}
DU^k = 0 \,, \ \omega_{pq} = 0
\,, \ \Theta =  \frac{\sqrt{2}\,L^{\prime}(u)}{L} \,,
\label{PW3bef}
\end{equation}
i.e., the acceleration four-vector and the
vorticity tensor are equal to
zero for this unit vector field $U^i$.
The corresponding shear tensor is
non-vanishing and can be written as a sum of two traceless tensors,
i.e.,
\begin{equation}
\sigma^k_i  = \frac{\Theta}{2}  \left(\frac13
\Delta_i^k - \delta_i^1 \delta^k_1 \right)
+ \frac{\beta^{\prime}}{\sqrt2} \left(\delta_i^2 \delta^k_2
{-} \delta_i^3 \delta^k_3 \right)\,.
\label{GW002}
\end{equation}
The gravitational
field equations for this case are
known to be compatible when the total stress-energy tensor is of the
null-type, i.e., it can be presented in the form $W \,k_ik_j$ with
$k_i$ a null
four-vector, $k_ik^i{=}0$. The analysis of the equations of the aether
motion shows that, when $U^i {=} \delta^i_t$, they can be satisfied
with some restrictions for the coupling parameters, but we refrain
from discussing details here.

\section{Conclusions}
\label{conc}

\subsection{On the interpretation of the coupling constants}
\label{interpretcpoup}

\subsubsection{Motivation}
\label{motivinterpret}

We have followed the rationale used for the
Einstein-aether theory, that for regions where quantum gravity
is not anymore dominant and Lorentz symmetry is already broken by
those quantum effects an Einstein-aether theory can naturally
appear \cite{J1}.
One
expects then that the Einstein-aether theory is a low energy
phenomenon of some fundamental quantum theory.
An Einstein-aether theory can be in action at the
inflationary period, giving rise to am
Einstein-scalar-aether theory \cite{solbar}.
A generic 2-form field,
like the Maxwell field, can appear at very high energy scales
giving rise to some form of an
Einstein-Maxwell-aether as discussed
by us (see also \cite{LIV31}).
Or it could be in operation after inflation
decays and the matter fields, such as the Maxwell field, make their
appearance.

\subsubsection{Coupling constants associated with a spontaneous
polarization-magnetization induced by a non-uniform aether motion}
\label{coupassoc}

In the Einstein-Maxwell-aether theory we have proposed,
the new cross-terms containing both the Maxwell tensor and the
covariant derivatives of the aether velocity four-vector, allows us
not only to give a formal interpretation of the new coupling
constants, but to propose ways of how one can try to estimate them
in the frameworks of the PPE and PPF formalisms. Such a work requires
detailed analysis and is beyond the scope here. Nevertheless, we would
like to expand our ideas in three examples.

According to Eqs.~(\ref{P4}) and (\ref{M4}), eight constants describe
the effects of spontaneous polarization and magnetization of the
matter or vacuum, which can appear due to an aether non-uniform
motion. The coupling constant $\pi_1$ introduces the polarization
produced by a pure acceleration of the aether. This parameter can pop
up in a static spherically symmetric system, since there the radial
component of the acceleration four-vector is non-vanishing, $DU_r \neq
0$. However, in static spherically symmetric systems the parameters
$\pi_2,\pi_3,\pi_4$ are hidden, since there are no shear, vorticity
and expansion in such spacetimes.  The parameter $\pi_2$ can appear
when the vector field has acceleration and expansion, $\Theta \neq 0$.
Similarly, a combination of acceleration and shear brings into the
open the parameter $\pi_3$. The combination of acceleration and
vorticity reveals the parameter $\pi_4$. Similar interpretation can be
done with the parameters $\mu_1, \mu_2, \mu_3$ (see Eq.~(\ref{M4})).
However, instead of the acceleration we have to use here the vorticity
tensor $\omega_{ik}$. The degeneracy with respect to the parameters
$\pi_2,\pi_3,\pi_4,\mu_1, \mu_2, \mu_3$ altogether can be removed, if
the spacetime contains a rotating object, like a neutron star and
thus is not spherically symmetric, or contains gravitational
waves propagating non-co-axially with respect to the aether motion.

\subsubsection{Coupling constants associated with optical activity
produced by an aether non-uniform motion}
\label{coupopti}

Optical activity is associated with the rotation of the polarization
of the electromagnetic waves propagating in a medium.  The presence of
optical activity amounts to the non-vanishing of the magneto-electric
coefficients tensor $\nu^{pm}$ (see, e.g., \cite{LL}).  According to
Eq.~(\ref{nu}) the couplings $\nu_1, ..., \nu_{12}$ describe the
optical activity of the matter or vacuum when the aether motion is
non-uniform. More precisely, the optical activity appears when the
aether is accelerated ($DU^i \neq 0$) or its velocity is characterized
by a non-vanishing vorticity tensor ($\omega_{ik} \neq 0$). Linear
effects in the vorticity tensor and in the acceleration four-vector
appear when $\nu_1$ and $\nu_2$ are non-vanishing, respectively.
Nonlinear effects appear when $\omega_{ik}\neq 0$ or $DU_k \neq 0$
and at least one of the two quantities $\Theta$ and $\sigma^{ik}$ is
not equal to zero. The removal of a degeneracy with respect to $\nu_2$
is possible for spherically symmetric objects. Other coupling
constants can appear in systems with rotating bodies or in systems
with gravitational waves with arbitrary direction of propagation. When
gravitational waves are present, the effects of optical activity are
similar to the ones described in \cite{BL02}.

\subsubsection{Coupling constants associated with dynamo-optical
effects and birefringence}
\label{coupdynamo}

Dynamo-optical effects are connected with the variation of the
dielectric and magnetic permittivity tensors in non-uniformly moving
media (see, e.g., \cite{LL}). When these permittivity tensors become
anisotropic, birefringence can take place, i.e., the phase
velocity of
electromagnetic waves is a function
of the wave polarization.  According to
Eqs.~(\ref{eps}) and (\ref{mu}) linear dynamo-optical effects induced
by an aether motion are connected to the presence of a shear tensor
and an expansion scalar, bringing into play the couplings $\alpha_1$,
$\alpha_6$, $\gamma_1$, and $\gamma_6$.  The other coupling constants
$\alpha_2,...,\alpha_{11}$ and $\gamma_2, ..., \gamma_{11}$ describe
quadratic and nonlinear cross-effects. One of the most interesting
application of these effects is the analysis of the phase and group
velocities of the electromagnetic waves propagating in the medium or
vacuum interacting with an aether non-uniform motion. Similar effects
caused by an interaction with curvature have been considered in
\cite{GW1,BL02,GW2,GW3,GW4,Num2,BBL12}.

\subsection{How can we reduce the number of coupling parameters
introduced phenomenologically?}
\label{reducenumber}

The Einstein-Maxwell-aether theory under consideration includes 79
independent coupling constants. It seems to be useful to reduce the
number of these parameters using some underlying symmetry, similarly
to what has been
done in nonminimal gravito-electric theories (see, e.g.,
\cite{BL05,BBL12}).  For instance, one can put
$C_2{=}C_3{=}C_4{=}0$ and keep only one constant $C_1\equiv C$, if we
admit that the squared contributions for the acceleration, shear,
vorticity and expansion are equivalent, see Eq.~(\ref{act6}). One can
also put $\pi_2{=}\pi_3{=}\pi_4$ and $\mu_2{=}\mu_3$ in order to
guarantee that the nonlinear terms enter in an equal manner into the
spontaneous polarization-magnetization tensor. This procedure can be
used for the permittivity tensors also. In this case the total number
of independent coupling constants can be reduced to 14, say.

\subsection{Outlook}
\label{outl}

The applications of this formalism to cosmology and astrophysics are
the next steps in the study of the Einstein-Maxwell-aether theory
proposed here. It is of interest to discuss Bianchi type I solutions in
this theory, as well as static and spherically symmetric solutions.
Of course, it will be important to make an analysis
of the Einstein-Maxwell-aether theory in the frameworks of PPN, PPF
and PPE formalisms.

\section*{Acknowledgments}

\noindent
AB is grateful to
the Russian Foundation for Basic Research (Grant No. 14-02-00598).
AB and JPSL are
grateful to FCT-Portugal for financial support
through the project
PTDC/FIS/098962/2008 and JPSL  is also
grateful to FCT-Portugal for financial support
through the project PEst-OE/FIS/UI0099/2011.

\appendix\section{Inclusion of
all the terms up to the fourth order in
an Einstein-Maxwell-aether theory and the choice
for the ansatz}
\label{appA}

\subsection{Extension of the Einstein-aether theory to include
all the terms up to fourth order in the derivatives}
\label{extensionm+n}

In the Secs.~\ref{inclusion} and \ref{action} we have  given
a motivation and the requirements to choose the ansatz
of the action functional as in Eq.~(\ref{E1})
an subsequent equations. Here we give the details
for such a choice.
We follow in part the structure of the action functional
for the pure Einstein-aether theory, see
Eq.~(\ref{1}), as discussed from
several points of view (see, e.g., \cite{J6} for a review),
and we
return to this question in order to justify further
generalizations that include the electromagnetic gauge vector field $A_i$
and the corresponding
gauge invariant Maxwell tensor $F_{ik}$.
According to the principles of effective field theories (see,
e.g., \cite{EFT1,EFT2,EFT3,EFT4,EFT5}) one can establish some
interrelations between the terms in the action functional and
differential operators of the first, second, and higher orders.

The tensor $F_{ik}$ is defined as
\begin{equation}
F_{ik}{=}\nabla_i A_k {-}\nabla_k A_i\,.
\label{maxwellA}
\end{equation}
The Maxwell tensor $F_{ik}$
seems to contain a covariant
derivative. However, due to the symmetry of the Christoffel symbols
$\Gamma^i_{km}{=}\Gamma^i_{mk}$ it can be rewritten using  partial
derivatives $F_{ik}{=}\partial_i A_k {-}\partial_k A_i$ only. In other
words, $F_{ik}$ contains neither metric coefficients, nor Christoffel
symbols, and thus this quantity does not change upon variation of the
action functional with respect to metric.
For this reason
we consider,
that the electromagnetic field has derivatives
independent of the derivatives involving the metric.
In a sense this means that
the electromagnetic field
introduces a scale parameter $l_{({\rm
em})}$ which is an independent scale. For instance,
in a cosmological setting, when we deal with, e.g.,
the cosmic microwave background
radiation, the electromagnetic derivatives, and
so
$l_{({\rm
em})}$, are of the order of the wavelength
of the radiation.

Now, in a theory of gravitation the covariant
derivative, $\nabla_i$, is the basic differential operator.
The
commutator $\nabla_i \nabla_k {-} \nabla_k \nabla_i$
of some vector field $U^m$ is known to
produce the Riemann tensor $R^m_{\ \ nik}$ according to the
relationship
\begin{equation}
(\nabla_i \nabla_k - \nabla_k \nabla_i) U^m = U^n R^m_{\ \ nik}\,.
\label{commut1}
\end{equation}
This means that, when we consider the Riemann tensor, the Ricci tensor
$R_{pq}{=}R^m_{\ \ pmq}$ and the Ricci scalar $R{=}R^p_p$, we deal, in
fact, with quantities of second order with respect to the covariant
derivative $\nabla_k$. Equivalently, these tensors are quantities up
to second order in the partial derivative of the metric.
This, in turn, means that
the metric
introduces a gravitational
scale parameter $l_{({\rm g})}$ which is another
independent scale. Such a scale can be
a cosmological distance, a radius of a star, or
any other relevant parameter.
In addition, concerning the terms of the
type $\nabla_a U_m$, we treat it as a
quantity of the first order in a metric derivative,
as this covariant
derivative contains a partial derivative of the metric and we suppose
that the gravity field alone makes the aether non-uniform.
Thus a derivative of the aether velocity also picks the
gravitational
scale parameter $l_{({\rm g})}$.
Note that the
covariant derivative of the Maxwell tensor,
$\nabla_m F_{ik}$
contains both types
of derivative, namely,
a second and first
order electromagnetic derivative
and a first order metric derivative.

The electromagnetic derivatives and the metric derivatives
are, in general, of different character.
For instance, in
a cosmological setting, when we deal with, e.g.,
the cosmic microwave background
radiation, the  electromagnetic derivative
is related to the wavelength
of the electromagnetic wave
and is of the order of 1 micrometer,
while at the same time the
metric derivative could be
of cosmological scale.

Based on these consideration below we use the following classification
for the scalar terms that can enter into the action functional: a
scalar term is of the type $(M,N)$ if it contains $M$th order metric
derivatives (i.e., it is of the $M$th order with respect to
$l^{-1}_{({\rm g})}$), and if it contains $N$th order electromagnetic
derivatives (i.e., it is of the $N$th order with respect to
$l^{-1}_{({\rm em})}$). This classification scheme is directly related
to the order $d$ scheme elaborated in \cite{LIV31} where $d=M+N$.
Our two parameter version of the classification of the Lagrangian
terms does not contradict this $d$ scheme and is, in fact, its
concretization. Within a given Lagrangian and action
with their corresponding coupling constants and terms,
our classification scheme is useful to pick up
the important terms in a given concrete
physical setting.

We want to display all the terms up to four orders
in the derivatives.
Thus,
let us discuss the structure of all the terms for which $M{+}N
\leq 4$. This means there is one type of zero-order terms: (0,0);
two types of the first order terms: (1,0) and (0,1); three types
of the second order terms: (2,0), (1,1) and (0,2); four types of
the third order terms: (3,0), (2,1), (1,2) and (0,3); five types
of the fourth order terms: (4,0), (3,1), (2,2), (1,3) and (0,4).

\begin{itemize}

\item
(0,0).
There is one term of this type involving the
aether velocity $U^i$. It is,
\begin{equation}
U^i\,U_i
\end{equation}
and is included
in the action functional of the standard Einstein-aether
theory. There are other two scalars, namely,
$
A_m A^m\quad{\rm and}\quad U_m A^m
$,
but, since they are not
gauge invariant, we
omit them.

\item
(1,0). The terms of this type are of the form
\begin{equation}
\alpha^{ik} \nabla_i
U_k\,.
\end{equation}
For the tensorial coefficients $\alpha^{ik}$,
constructed using the
metric $g_{ik}$, the Kronecker tensors
($\delta^i_k$, $\delta^{ik}_{ab}$ and higher order Kronecker tensors),
the Levi-Civita tensor $\epsilon^{ikab}$, and the unit vector field
$U^k$, there is only one appropriate scalar of the
type (1,0), namely, $\alpha \Theta$, where $\Theta{=}\nabla_kU^k$ is
the expansion scalar, and $\alpha$ is a coupling constant
introduced phenomenologically.

\item
(0,1). There are no gauge-invariant scalars of the type (0,1) that
would contain the Maxwell tensor in combination with the metric
$g_{ik}$, the Kronecker tensors ($\delta^i_k$,
$\delta^{ik}_{ab}$ and higher order Kronecker tensors), the
Levi-Civita tensor $\epsilon^{ikab}$, and the unit vector field $U^k$.

\item
(2,0). The type (2,0) is exhausted by the terms
\begin{equation}
R\,,
\end{equation}
and
\begin{equation}
K^{abmn}
\nabla_a U_m \nabla_b U_n\,,
\end{equation}
which enter in the action functional
(\ref{1}) of the Einstein-aether theory.

There are other terms, but these can be absorbed or discarded.
Indeed, terms with second-order covariant derivatives $ {\cal
K}^{ikl}\nabla_i \nabla_k U_l$, in which $ {\cal K}^{ikl}$
contains the metric $g_{ik}$, the Kronecker tensors ($\delta^i_k$,
$\delta^{ik}_{ab}$ and higher order Kronecker tensors), the
Levi-Civita tensor $\epsilon^{ikab}$, and the unit vector field
$U^k$, can be rewritten as follows
$
{\cal K}^{ikl}\nabla_i \nabla_k U_l =\nabla_i \left[{\cal K}^{ikl}
\nabla_k U_l\right] - (\nabla_k U_l) \nabla_i \left({\cal
K}^{ikl}\right)\,.
$
Since the
metric $g_{ik}$, the Kronecker tensors
($\delta^i_k$, $\delta^{ik}_{ab}$ and higher order Kronecker tensors),
and the Levi-Civita tensor $\epsilon^{ikab}$,
are covariantly constant tensors, i.e., $\nabla_l g_{ik}=0$,
$\nabla_l \delta^i_k=0$, $\nabla_l \delta^{ik}_{mn}=0$, $\nabla_l
\epsilon^{ikmn}=0$, we obtain from the above mentioned term,
$
{\cal K}^{ikl}\nabla_i \nabla_k U_l = \nabla_i \left[{\cal
K}^{ikl} \nabla_k U_l\right] - (\nabla_k U_l) (\nabla_i U^j)
\frac{\partial {\cal K}^{ikl}}{\partial U^j} \,.
$
The first term in the right-hand side of this relationship is a
perfect four-divergence, which can be omitted, and
the second term can be
included into $K^{abmn} \nabla_a U_m \nabla_b U_n$ by redefinition of
the tensor $K^{abmn}$.

As for the nonminimal term $R_{ik}U^i U^k$ it can also be absorbed and
discarded. Using (\ref{commut1}), $R_{ik}U^i U^k$ can be rewritten as
$
R_{ik}U^i U^k = \nabla_i \left[U^k \nabla_k U^i- U^i\nabla_k U^k
\right]+ \left(\nabla_i U^i \right)\left(\nabla_k U^k \right) -
\left(\nabla_m U^k \right)\left(\nabla_k U^m \right) \,.
$
The first term in this relationship is a perfect four-divergence,
and the other terms can be included in the construction of the
Jacobson's type term $K^{abmn} (\nabla_a U_m)(\nabla_b U_n)$. Here
and below we use the parentheses in the expressions of the form
$(\nabla_a U_m) {\cal T}$ just to indicate that the covariant
derivative operator acts on $U_m$ only.

\item
(1,1). The gauge-invariant terms of the type (1,1) can be listed using the
representation
\begin{equation}
A^{mnpq}F_{pq} \nabla_{m}U_n \,,
\label{3091}
\end{equation}
where the tensor coefficients $A^{mnpq}$ are constructed using the
metric $g_{ik}$, the covariant constant Kronecker tensors
($\delta^i_k$, $\delta^{ik}_{ab}$ and higher order Kronecker tensors),
the Levi-Civita tensor $\epsilon^{ikab}$, and the unit vector field
$U^k$.

There are also terms of the type
${\cal A}^{mpq} \nabla_{m}F_{pq}$. However, these
can be reduced to the terms given in Eq.~(\ref{3091})
using the relationships
$
{\cal A}^{mpq} \nabla_{m}F_{pq} = \nabla_{m} \left[{\cal A}^{mpq}
F_{pq} \right] - F_{pq} (\nabla_{m}U^j) \frac{\partial {\cal
A}^{mpq}}{\partial U^j} \,,
$
with the corresponding redefinition of the quantity $A^{mnpq}$.

\item
(0,2). The representatives of the type (0,2) are given by
$F_{mn}F^{mn}$ and $F_{mn}U^n F^{ml}U_l$. Generically, such
terms can be described as
\begin{equation}
{\cal C}^{ikmn}_{(2)} F_{ik} F_{mn} \,,
\label{307}
\end{equation}
where ${\cal C}^{ikmn}_{(2)}$ is called the linear response tensor.
The subscript $(2)$ indicates here that this term is quadratic in the
Maxwell tensor $F_{ik}$.

\item (3,0). The type (3,0) includes terms of three subtypes:
\begin{equation}
Z^{ikmnls}_{(1)}  (\nabla_i U_k) (\nabla_m U_n)  (\nabla_l U_s)  \,,
\label{301}
\end{equation}
\begin{equation}
Z^{imnls}_{(2)}  (\nabla_i \nabla_m U_n)  (\nabla_l U_s) ,
\label{302}
\end{equation}
\begin{equation}
Z^{ikmnls}_{(3)}  R_{ikmn}\nabla_l U_s  \,.
\label{303}
\end{equation}
There are also terms of the type $Z^{imls}_{(4)}\nabla_i \nabla_m
\nabla_l U_s$. However, these can be transformed into a
combination of the terms given in Eqs.~(\ref{301}) and (\ref{302})
by the procedure described for the (2,0) type
terms. In addition, terms of the type
$Z^{ikmnl}_{(5)}\nabla_l R_{ikmn}$ can be expressed as the terms
in Eq.~(\ref{303}) using integration by parts, namely
$
Z^{ikmnl}_{(5)}\nabla_l R_{ikmn} = \nabla_l \left[Z^{ikmnl}_{(5)}
R_{ikmn}\right] - R_{ikmn} (\nabla_l U_j) \frac{\partial
Z^{ikmnl}_{(5)}}{\partial U^j} \,.
$

\item
(2,1). The list of independent terms of the type (2,1) is:
\begin{equation}
B^{mnlspq} F_{pq}(\nabla_m U_n) (\nabla_l U_s)  \,,
\label{list111}
\end{equation}
\begin{equation}
{\cal B}^{mlspq} F_{pq} \nabla_m \nabla_l U_s  \,,
\label{list112}
\end{equation}
\begin{equation}
Q^{ikmnpq}R_{ikmn} F_{pq} = Q R^{ik} U_k F_{im}U^m \,.
\label{list113}
\end{equation}
There are also terms of the type ${\cal Q}^{iklpq}_{(1)} (\nabla_l
F_{pq}) (\nabla_i U_k)$, but due to the relationships
$
{\cal Q}^{iklpq}_{(1)} (\nabla_l F_{pq}) (\nabla_i U_k) = \nabla_l
\left[{\cal Q}^{iklpq}_{(1)} F_{pq} (\nabla_i U_k) \right] {-}
F_{pq} (\nabla_l U^j)  (\nabla_i U_k) \frac{\partial {\cal
Q}^{iklpq}_{(1)}}{\partial U^j} {-} {\cal Q}^{iklpq}_{(1)} F_{pq}
(\nabla_l \nabla_i U_k)
$
these terms
can be reduced to the terms given in
Eqs.~(\ref{list111}) and (\ref{list112}).
Similarly, terms in the second derivative of the Maxwell tensor,
i.e.,
${\cal Q}^{ilpq}_{(2)} (\nabla_i\nabla_l F_{pq})$ can be transformed
into terms of the type ${\cal Q}^{iklpq}_{(1)} (\nabla_l F_{pq})
(\nabla_i U_k)$ and then be reduced again to
the terms given in Eqs.~(\ref{list111}) and
(\ref{list112}).

\item
(1,2). The terms of the type (1,2) can be specified as:
\begin{equation}
X^{mnikpq} (\nabla_m U_n) F_{ik} F_{pq}  \,.
\label{list2}
\end{equation}
There are also terms of the type $X^{mikpq}_{(1)} F_{ik} (\nabla_m
F_{pq})$ which, again, can be transformed into the terms given in
Eq.~(\ref{list2}) by integration by parts.

\item
(0,3). The terms of the type (0,3)
can be written as
\begin{equation}
{\cal C}^{ikmnls}_{(3)} F_{ik} F_{mn} F_{ls} \,,
\label{308}
\end{equation}
where ${\cal C}^{ikmn}_{(3)}$ is a second-order
response tensor.

\item
(4,0). We divide the type (4,0) into three subtypes. The first one
contains various quadratic combinations of the Ricci scalar, Ricci
and Riemann
tensors, and the unit four-vector $U^j$, and can be
written in an abbreviated form as
\begin{equation}
{\cal Z}^{ikmnlspq}_{(1)}  R_{ikmn} R_{lspq}  \,.
\label{304}
\end{equation}
The second subtype consists of combinations of the Ricci scalar,
Ricci and Riemann tensors multiplied by covariant derivatives of the
unit four-vector,  and can be
written as two terms, namely,
\begin{equation}
{\cal Z}^{ikmnlspq}_{(2)}  R_{ikmn} (\nabla_lU_s) (\nabla_pU_q)
\,,  \quad {\cal Z}^{ikmnlpq}_{(3)}  R_{ikmn} (\nabla_l
\nabla_pU_q) \,. \label{305}
\end{equation}
Again, all other terms, which contain $\nabla_p \nabla_q R_{ikmn}$
and
$(\nabla_p U^j)(\nabla_q R_{ikmn})$ can be transformed into
combinations of the already listed terms in Eq.~(\ref{305}).

The third subtype does not include the Riemann tensor, contains
combinations of the covariant derivatives of the unit
four-vector, and can be
written as three terms, namely,
$$
{\cal Z}^{ikmnlspq}_{(4)}(\nabla_i U_k) (\nabla_m U_n)
(\nabla_lU_s) (\nabla_pU_q) \,,
$$
\begin{equation}
{\cal Z}^{imnlspq}_{(5)}(\nabla_i \nabla_m U_n) (\nabla_lU_s)
(\nabla_pU_q) \,, \quad {\cal Z}^{imnlpq}_{(6)}(\nabla_i \nabla_m
U_n) (\nabla_l \nabla_pU_q) \,. \label{306}
\end{equation}
 Similarly, the terms that contain
$\nabla_p \nabla_q \nabla_a U_j$ and $\nabla_p
\nabla_q \nabla_a \nabla_b U_j$ can be transformed into
combinations of the already listed terms in Eq.~(\ref{306}).

Let us stress, that in fact we can consider in
Eq.~(\ref{306}) only the
symmetrized terms $\nabla_{(i} \nabla_{m)} U_n$, since, according to
(\ref{commut1}), its skew-symmetric part $\nabla_{[i} \nabla_{m]} U_n$
can be expressed using the Riemann tensor, and the corresponding
scalar,
$
{\cal Z}^{imnlspq}_{(5)}(\nabla_{[i} \nabla_{m]} U_n)
(\nabla_lU_s) (\nabla_pU_q) = \frac12 {\cal Z}^{imnlspq}_{(5)} U^s
R_{nsim} (\nabla_lU_s) (\nabla_pU_q) \,,
$
can be reduced to the terms given in
Eq.~(\ref{305}). We are using the
standard symbols for symmetrization ${\cal T}_{(ik)}{=}\frac12
\left[{\cal T}_{(ik)}{+}{\cal T}_{(ki)}\right]$, and skew-symmetrization
${\cal T}_{[ik]}{=}\frac12 \left[{\cal T}_{(ik)}{-}{\cal T}_{(ki)}\right]$.

\item
(3,1). The terms of the type (3,1)
can be written by a simple extension of the
nomenclature used for the terms of the
type (3,0), i.e.,
\begin{equation}
{\tilde Z}^{ikmnlspq}_{(1)} F_{pq} (\nabla_i U_k) (\nabla_m U_n)
(\nabla_l U_s)  \,,
\label{311}
\end{equation}
\begin{equation}
{\tilde Z}^{imnlspq}_{(2)}  F_{pq} (\nabla_i
\nabla_m U_n)  (\nabla_l U_s) \,,
\label{312}
\end{equation}
\begin{equation}
{\tilde Z}^{ikmnpq}_{(3)} F_{pq} \nabla_i \nabla_k \nabla_m U_n  \,,
\label{313}
\end{equation}
\begin{equation}
{\tilde Z}^{ikmnlspq}_{(4)} F_{pq} R_{ikmn}\nabla_l U_s  \,,
\label{314}
\end{equation}
\begin{equation}
{\tilde Z}^{ikmnlpq}_{(5)} F_{pq} \nabla_l R_{ikmn}  \,.
\label{315}
\end{equation}
The terms in the covariant derivative of the Maxwell tensor,
$\nabla_j F_{pq}$, can be reduced to the listed terms
in Eqs.~(\ref{311})-(\ref{315}) by an
integration by parts, not being necessary to repeat
the procedure here.

\item (2,2). The type (2,2) is relevant in our considerations.
There are four subtypes.

The first subtype contains the covariant derivatives of the unit
vector field, but does not include the nonminimal terms
constructed using the Ricci scalar, and the Ricci and Riemann
tensors.  It has two terms, namely,
\begin{equation}
Y^{mnlsikpq} F_{ik} F_{pq} (\nabla_m U_n) (\nabla_l U_s)  \,,
\quad {\cal Y}^{mlsikpq}F_{ik} F_{pq} (\nabla_m \nabla_l U_s)\,.
\label{list3}
\end{equation}
The second subtype contains only nonminimal terms
\cite{How,DH,BLZ2010} (see also
\cite{BL05,BBL12,BL02,GW1,GW2,GW3,GW4,Num2}),
it does not
contain the unit vector field $U^l$. The independent terms are
three, namely,
\begin{equation}
R F_{ik} F^{ik} \,, \quad R^{ik} F_{im} F_{k}^{\ m} \,,
\quad R^{ikmn} F_{ik} F_{mn}
\label{list4}
\end{equation}
There are other nonminimal terms, i.e., $R^{ikmn} F^{*}_{ik}
F^{*}_{mn}$, $^{*}R^{ikmn} F_{ik} F^{*}_{mn}$, $^{*}R^{ikmn}
F^{*}_{ik} F_{mn}$, $R^{*ikmn} F^{*}_{ik} F_{mn}$, and so on, where an
asterisk means we are taking the dual of the respective tensor with
the Levi-Civita tensor. However, these terms can be reduced to a
combination of the terms given in Eq.~(\ref{list4}).

The third subtype
includes independent combinations of the following scalars,
$R F_{im}U^m F^{in} U_n$, $\,R_{pq}U^p U^q F_{ik} F^{ik}$,
$\,R_{pq}U^p U^q F_{im}U^m F^{in} U_n$, $\,R_{pq}U^q
F_{im}U^m F^{ip}$,
$\,R^{ikmn} U_k U_n F_{ip} F_{m}^{\ \ p}$,
$\,R^{ikmn} U_k F_{ip} U^p F_{mn}$
$\, R^{ikmn} U_k U_n F_{ip}U^p F_{mq} U^q$,
and their analogs containing the pairs of dual quantities $R^{*ikmn}$
with $F^{*}_{pq}$ and $^{*}R^{ikmn}$ with $F^{*}_{pq}$.
Generically, all
these terms can be written as
\begin{equation}
S^{ikmnlspq} R_{ikmn} F_{ls} F_{pq} \,,
\label{list6}
\end{equation}
They are extensions of the nonminimal terms.

The fourth subtype
includes the irreducible terms which are quadratic in the
covariant derivatives of the Maxwell tensor, namely,
\begin{equation}
G^{pikqmn}(\nabla_p F_{ik}) (\nabla_q F_{mn}) \,.
\label{list7}
\end{equation}
There are other terms that could be included.
However, the terms that
contain first covariant derivatives of the Maxwell
tensor $G^{ikpmnqj}_{(1)}F_{ik}(\nabla_p F_{mn}) (\nabla_q U_{j})$ can
be reduced to a combination of the terms
given in Eq.~(\ref{list3}). In addition, the scalars
in the second covariant derivative of the Maxwell tensor
$G^{ikpqmn}_{(2)}F_{ik}(\nabla_p \nabla_q F_{mn})$ can also be
transformed into terms of the type (2,2) already listed above.

\item
(1,3). Similarly to the case (0,3) one obtains that all the terms
of this subtype can be written as
\begin{equation}
{\cal C}^{ikmnlspq}_{(5)} F_{ik} F_{mn} F_{ls} \nabla_{p}U_q \,.
\label{131}
\end{equation}
The terms containing $\nabla_i F_{pq}$ again can be reduced to
the scalars of the type given in Eq.~(\ref{131}).

\item
(0,4). Similarly to the cases (0,2) and (0,3) one obtains the terms
\begin{equation}
{\cal C}^{ikmnlspq}_{(4)} F_{ik} F_{mn} F_{ls} F_{pq}\,,
\label{309}
\end{equation}
where the tensor ${\cal C}^{ikmnlspq}_{(4)}$ describes a
nonlinear electromagnetic response of the third order.

\end{itemize}

\subsection{Remarks}
\label{2remarks}

Some remarks related to our classification $M+N \leq 4$
scheme are in order.

\noindent
{(i)}
Following the study of the dynamical evolution of a scalar
field $\phi$ in the primordial universe \cite{EFT2},
in the framework of effective field
theory, one has that the derivative of a scalar field $\nabla_k
\phi {=}\partial_k \phi$ is considered as a
metric derivative, i.e., it is a
quantity of the order
$l^{-1}_{({\rm g})}$, since the variations of the scalar field are
produced by the dynamics of the gravitational
field. Thus, terms
of the type $R_{ik} \nabla^k \phi \nabla^i \phi$
and $(g^{pq}\nabla_p
\nabla_q \phi) \, (\nabla^k \phi) (\nabla_k \phi)$
that appear in the Lagrangian
presented in
\cite{EFT2} are metric derivatives of
the fourth order.
There is a correspondence to our case.
First, $\nabla_m U_n$ is a metric derivative of
first order, i.e., of the order $l^{-1}_{({\rm
g})}$. Second,
terms of the type $R^{ik}\nabla_iU_m \nabla_k U^m$ and
$(U^p\nabla_p \nabla_q U^q) \, (\nabla^k U_l) (\nabla_k U^l)$
are then metric derivatives of
the fourth order, i.e., of the type (4,0) in
our classification.

\noindent
{(ii)} Following \cite{DH} the electromagnetic derivatives
and the metric derivatives are independent,
i.e., the parameters $l_{({\rm g})}$ and
$l_{({\rm em})}$ to be independent. This means, for instance, that the
terms of the type (2,2) (see Eq.~(\ref{list3})) can be of the same order
of magnitude as the terms of the type (1,1), when the wavelength of an
electromagnetic wave $ \lambda_{({\rm em})}\simeq
l_{({\rm em})}$ is of
the order of $l_{({\rm g})}\frac{Y}{A}$, where $Y$ and $A$ are the
typical values of the components of the tensors $Y^{mnlsikpq}$ and
$A^{mnpq}$, respectively. Similarly, there are cases when the terms of
the type (2,2) can be considered as leading order terms in comparison
with, e.g.,  terms of the type (4,0).  There are also
special cases, when $l_{({\rm em})}$ and $l_{({\rm g})}$ are of the
same order, and we should consider terms of the type (2,2) to be of
the same order of magnitude, as terms, e.g., of the type (4,0) and
(3,1). That is why we listed all the terms of the type $(M,N)$, for
which $M+N \leq 4$.

\noindent
{(iii)}
We note that a
general formulation of the Einstein-Maxwell-aether theory does not
allow the explicit introduction of the parameter $l_{({\rm g})}$
and the definition of the corresponding dimensionless coupling
constants. However, this becomes possible, when one deals with
applications of the theory to cosmology and astrophysics. For
instance, in \cite{BLZ2010} studying nonminimal traversable
electric wormholes we have introduced three parameters with the
dimension of length, namely the gravitational $r_{({\rm M})}$ and
the electric $r_{({\rm Q})}$ radii, related to the mass $M$ and
charge $Q$ of the object, respectively, as well as, the nonminimal
radius $r_{({\rm q})}$ connected with the nonminimal coupling
parameter $|q_1|$. Dimensionless parameters $\frac{r_{({\rm
q})}}{r_{({\rm M})}}$ and $\frac{r_{({\rm Q})}}{r_{({\rm M})}}$
became then the guiding parameters in the analysis of the wormhole
solution. We expect that in applications of the
Einstein-Maxwell-aether theory the introduction of coupling
parameters will appear naturally.

\noindent
{(iv)}
In addition, from the point of view of dimensional units, the
coefficients $A^{mnpq}$ and $B^{mnlspq}$, $X^{mnikpq}$ and
$Y^{mnlsikpq}$ (see Eq.~(\ref{M2})) differ from each others by powers
in units of length.  In a generic formulation there is no interest in
introducing multipliers to provide the same dimensionality for these
tensorial objects. On the other hand, a
units redefinition of the coupling parameters
will perhaps be of interest when one deals with applications of the theory.

\subsection{The ansatz}
\label{ansa}

\subsubsection{Requirements for the ansatz}
\label{ansareq}
We impose now three requirements that our theory should satisfy.

\noindent
{(a)} The electrodynamics of the theory must be linear in the Maxwell
tensor $F_{ik}$ and of second order in the partial derivatives of the
electromagnetic potential four-vector $A_i$.  These requirements imply
that the terms of the type (0,3), (1,3), (0,4), given in
Eqs.~(\ref{308}), (\ref{131}), and (\ref{309}), respectively, and the
term given in Eq.~(\ref{list7}) quadratic in the derivative of the
Maxwell tensor of the type (2,2) are not present in the theory.

\noindent {(b)} The dynamical equations for the unit vector field
$U^i$ are considered to be a set of quasilinear equations of
second order in their partial derivatives.
This requirement
implies that the terms of the type (3,0) given in
Eqs.~(\ref{301}) and (\ref{302}),
of the type
(4,0) given in
Eq.~(\ref{306}),
and of the type
(3,1) given in
Eqs.~(\ref{311}), (\ref{312}), and
(\ref{313}), are not present in
the theory.
A note is in order:
According to the standard terminology in
mathematical physics quasilinear
means that the equations can be nonlinear in
the four-vector $U^i$ itself, nonlinear in the first covariant
derivative $\nabla_i U_k$, but the second partial
derivatives $\partial_{i} \partial_k U_s$ enters the equations
linearly with tensorial coefficients that can depend on $U^i$ and
$F_{mn}$, but can not contain $\nabla_i U_k$.

\noindent
{(c)} The equations for the gravitational
field are considered to be
equations of
second order in the partial derivatives of the metric
(similarly to the standard Einstein's and Einstein-aether
theories).
This requirement
implies that the terms of the type (3,0) given in
Eq.~(\ref{303}),
of the type
(4,0) given in
Eqs.~(\ref{304}) and (\ref{305}),
and of the type
(3,1) given in
Eqs.~(\ref{314}), and
(\ref{315}), are not present in
the theory.
All the other terms are included into the action functional of the
theory we propose, see next section.

This set of requirements (a), (b), and (c)
can be reformulated as the assumption that the discarded
terms have coefficients, phenomenologically introduced, that are small
enough in comparison with the non-discarded coupling constants.

\subsubsection{The ansatz}
\label{ansa2}
With these requirements, the ansatz for
the Lagarangian and the action can then be given as in
Eq.~(\ref{E1}) and subsequent equations.

\section{Reconstruction of the electrodynamic
tensors $X^{lsikmn}$ and $Y^{ablsikmn}$
in terms of electrodynamic constants and
spacetime tensors}
\label{App:AppendixB}

We recall that in a medium moving with
velocity $U^i$ the currentless equations
of electrodynamics can be rewritten as
the four Maxwell equations and two
constitutive equations, as stated in
Sect.~\ref{electroeq}. The
four Maxwell equations are the Gauss law
\begin{equation}
\Delta_k^m \nabla_m {\cal D}^k =  \omega_k {\cal H}^k \,,
\label{EHDB51}
\end{equation}
the law of the magnetic flux conservation
\begin{equation}
\Delta_k^m \nabla_m B^k = - \omega_k E^k \,,
\label{EHDB4}
\end{equation}
the Amp\`ere law
\begin{equation}
\Delta^{ik} D {\cal D}_k - \eta^{ikm} \nabla_k {\cal H}_m = - 2
\Delta^i_k {\cal H}_m \omega^{*km} + \left(\sigma^{ik} {-}
\omega^{ik} {-} \frac23 \Theta \Delta^{ik} \right) {\cal D}_k \,,
\label{EHDB52}
\end{equation}
and the Faraday law,
\begin{equation}
\Delta^{ik} DB_k + \eta^{ikm} \nabla_k E_m = 2 \Delta^i_k E_m
\omega^{*km} + \left(\sigma^{ik} {-} \omega^{ik} {-} \frac23
\Theta \Delta^{ik} \right)B_k \,. \label{EHDB5}
\end{equation}
The constitutive equations are
\begin{equation}
{\cal D}^i = {\cal P}^i + \varepsilon^i_k E^k - \nu^{ki}B_k \,,
\quad  {\cal H}^i = {\cal M}^i + (\mu^{{-}1})^{ik} B_k +
\nu^{ik}E_k \,. \label{EHDB3}
\end{equation}
We used here the standard definition
$\omega^i \equiv {-} \epsilon^{ikmn} U_n \nabla_k U_m$ for the local
angular rotation velocity of the medium.

The reconstruction started in Sect.~\ref{xandy}
of the quantities $X^{mnabpq}$ and
$Y^{mnlsabpq}$ yields, respectively,
$$
X^{lsikmn} =
\frac12 \left(\alpha_1 {-} \frac13 \alpha_6 \right)
\Delta^{ls}\left(g^{ikmn}-\Delta^{ikmn} \right) +
\frac14 \alpha_6 U_p U_q \left[g^{iklp}g^{mnsq} +
g^{mnlp}g^{iksq} \right]+
$$
\begin{equation}
+ \frac12 \left(\gamma_1 {-} \frac13 \gamma_6 \right)
\Delta^{ls}\Delta^{ikmn} - \frac12 \gamma_6 \
\eta^{ik(l} \eta^{s)mn} - \nu_2 U^l \left\{\Delta^{iks[m}U^{n]}
+  \Delta^{mns[i}U^{k]}\right\}
\,,
\label{X}
\end{equation}
$$
Y^{ablsikmn} =
\frac12 \left(g^{ikmn}{-}\Delta^{ikmn} \right) \left[
\alpha_2 U^a U^l \Delta^{bs} +
\left(\alpha_3 {-} \frac13 \alpha_4 {+}\frac19
\alpha_9\right)\Delta^{ab} \Delta^{ls}
+ \alpha_4 \Delta^{a(l}\Delta^{s)b} +
\frac12 \alpha_5 \Delta^{abls}
\right] {-}
$$
$$
{-} \frac14 \alpha_7 \left\{
U_pU_q \left[\Delta^{ab} g^{ikp(l} g^{s)qmn}
{+} \Delta^{ls} g^{ikp(a} g^{b)qmn} \right]
{+}\frac23 \Delta^{ab} \Delta^{ls}
\left(g^{ikmn}{-}\Delta^{ikmn} \right)\right\} {-}
$$
$$
- \frac12 \alpha_8 U^a U^l U_p U_q \ g^{ikp(b} g^{s)qmn}  +
$$
$$
+ \frac12 \alpha_9 U_pU_q  \left\{
\frac13 \ \left[\Delta^{ab}g^{ikp(l} g^{s)qmn} {+}
\Delta^{ls}g^{ikp(a} g^{b)qmn}\right] {-}
\frac12 \left[g^{ikp(a} \Delta^{b)(l}g^{s) qmn} +
g^{ikp(l} \Delta^{s)(a}g^{b) qmn} \right]
\right\} {+}
$$
$$
{+} \frac18 \alpha_{10} U_p \left\{
g^{ikbp} U^{[m} \Delta^{n]als} - g^{ikap} U^{[m} \Delta^{n]bls}
+ g^{mnbp}U^{[i} \Delta^{k]als} - g^{mnap} U^{[i} \Delta^{k]bls} +
\right.
$$
$$
\left.
+ g^{iksp} U^{[m} \Delta^{n]lab} - g^{iklp} U^{[m} \Delta^{n]sab}
+ g^{mnsp} U^{[i} \Delta^{k]lab} - g^{mnlp} U^{[i} \Delta^{k]sab}
\right\} -
$$
$$
- \frac18 \alpha_{11}U_p \left\{
g^{ikap}U^{[m}\Delta^{n]bls} +
g^{ikbp}U^{[m}\Delta^{n]als} +
g^{mnap}U^{[i}\Delta^{k]bls}+ g^{mnbp}U^{[i}\Delta^{k]als} +
\right.
$$
$$
\left.
{+} g^{iklp}U^{[m}\Delta^{n]sab} +
g^{iksp}U^{[m}\Delta^{n]lab} +
g^{mnlp}U^{[i}\Delta^{k]sab}+ g^{mnsp}U^{[i}\Delta^{k]lab}
\right\} +
$$
$$
+\frac12 \Delta^{ikmn}\left[
\gamma_2 U^a U^l \Delta^{bs} +
\left(\gamma_3 {-} \frac13 \gamma_4 \right)\Delta^{ab} \Delta^{ls}
+ \gamma_4 \Delta^{a(l}\Delta^{s)b} + \frac12 \gamma_5 \Delta^{abls}
\right] -
$$
$$
-\frac14 \gamma_7 \left\{
\Delta^{ab} \eta^{ik(l}\eta^{s)mn} +
\Delta^{ls} \eta^{ik(a}\eta^{b)mn} +
\frac23 \Delta^{ab}\Delta^{ls} \Delta^{ikmn}
\right\} -
$$
$$
- \frac14 \gamma_8 U^a U^l \left(\eta^{ikb}\eta^{mns} +
\eta^{mnb}\eta^{iks} \right) + \frac12 \gamma_9 \left\{\frac13
\left[\Delta^{ab} \eta^{ik(l}\eta^{s)mn} + \Delta^{ls}
\eta^{ik(a}\eta^{b)mn} \right] - \right.
$$
$$
\left.
- \frac14 \left[\eta^{ikb}\eta^{mn(l}
\Delta^{s)a} + \eta^{ika}\eta^{mn(l}\Delta^{s)b}
+\eta^{mnb}\eta^{ik(l}\Delta^{s)a} +
\eta^{mna}\eta^{ik(l}\Delta^{s)b}
\right] \right\} +
$$
$$
+ \frac{1}{16} \gamma_{10} \left\{
\eta^{iks}\eta^{mn[a}\Delta^{b]l} - \eta^{ikl}\eta^{mn[a}\Delta^{b]s}
+ \eta^{mns}\eta^{ik[a}\Delta^{b]l} - \eta^{mnl}\eta^{ik[a}\Delta^{b]s}
+
\right.
$$
$$
\left.
+\eta^{ikb}\eta^{mn[l}\Delta^{s]a} - \eta^{ika}\eta^{mn[l}\Delta^{s]b}
+ \eta^{mnb}\eta^{ik[l}\Delta^{s]a} - \eta^{mna}\eta^{ik[l}\Delta^{s]b}
\right\} -
$$
$$
-\frac18 \gamma_{11} \left\{
\eta^{ik[a} \Delta^{b](l}
\eta^{s)mn} + \eta^{mn[a} \Delta^{b](l} \eta^{s)ik}
+\eta^{ik[a} \Delta^{b](l}
\eta^{s)mn} + \eta^{mn[a} \Delta^{b](l} \eta^{s)ik}
\right\} -
$$
$$
- \frac12\nu_4 \left\{\Delta^{ab} U^l
\left[\Delta^{iks[m}U^{n]}+ \Delta^{mns[i}U^{k]}\right]
+ \Delta^{ls} U^a \left[\Delta^{ikb[m}U^{n]}
+ \Delta^{mnb[i}U^{k]}\right]
\right\}+
$$
$$
+ \frac18\left(\nu_5{-}\nu_6 \right) \left\{
\Delta^a_p \Delta^b_q
\left[U^{[m}\epsilon^{n]pq(l} \eta^{s)ik} +
U^{[i}\epsilon^{k]pq(l} \eta^{s)mn} \right] +
\Delta^l_p \Delta^s_q
\left[U^{[m}\epsilon^{n]pq(a} \eta^{b)ik}
+ U^{[i}\epsilon^{k]pq(a} \eta^{b)mn} \right]
\right\} +
$$
$$
+ \frac{1}{32}\left(\nu_7
{+}\nu_8 \right)\Delta^{abjt}\Delta^{lspq}\left[
\eta^{ik}_{\ \ p}U^{[m}
\epsilon^{n]}_{\ \ qjt} +
\eta^{mn}_{\ \ \ p}U^{[i}\epsilon^{k]}_{\ \ qjt} +
\eta^{ik}_{\ \ j}U^{[m}
\epsilon^{n]}_{\ \ tpq} +
\eta^{mn}_{\ \ \ j}U^{[i}\epsilon^{k]}_{\ \ tpq}
\right] +
$$
$$
+ \frac18 \nu_9 \left\{
U^a \left[
\eta^{ikb}U^{[m}\eta^{n]ls} +
\eta^{mnb}U^{[i}\eta^{k]ls} - \Delta^{ikls}U^{[m}\Delta^{n]b}
- \Delta^{mnls}U^{[i}\Delta^{k]b}\right] + \right.
$$
$$
\left.
+ U^l \left[ \eta^{iks}U^{[m}\eta^{n]ab} +
\eta^{mns}U^{[i}\eta^{k]ab} - \Delta^{ikab}U^{[m}\Delta^{n]s}
- \Delta^{mnab}U^{[i}\Delta^{k]s}\right]\right\} {+}
$$
$$
+ \frac18 \nu_{10} \left\{
U^a \left[
\eta^{ikb}U^{[m}\eta^{n]ls} +
\eta^{mnb}U^{[i}\eta^{k]ls} + \Delta^{ikls}U^{[m}\Delta^{n]b}
+ \Delta^{mnls}U^{[i}\Delta^{k]b}\right] + \right.
$$
$$
\left.
+ U^l \left[ \eta^{iks}U^{[m}\eta^{n]ab} +
\eta^{mns}U^{[i}\eta^{k]ab} + \Delta^{ikab}U^{[m}\Delta^{n]s}
+ \Delta^{mnab}U^{[i}\Delta^{k]s}\right]
\right\} +
$$
$$
+ \frac18 \nu_{11} \left\{
U^a\left[
\eta^{iks}U^{[m}\eta^{n]lb} +
\eta^{ikl}U^{[m}\eta^{n]sb} +
\eta^{mns}U^{[i}\eta^{k]lb} + \eta^{mnl}U^{[i}\eta^{k]sb}
-
\right.\right.
$$
$$
\left.\left.
- \Delta^{iklb}U^{[m}\Delta^{n]s} -
\Delta^{iksb}U^{[m}\Delta^{n]l} -
\Delta^{mnlb}U^{[i}\Delta^{k]s} - \Delta^{mnsb}U^{[i}\Delta^{k]l}
\right] + \right.
$$
$$
\left.
+
U^l\left[
\eta^{ikb}U^{[m}\eta^{n]as} +
\eta^{ika}U^{[m}\eta^{n]bs} +
\eta^{mnb}U^{[i}\eta^{k]as} + \eta^{mna}U^{[i}\eta^{k]bs}
-
\right.\right.
$$
$$
\left.\left.
- \Delta^{ikas}U^{[m}\Delta^{n]b} -
\Delta^{ikbs}U^{[m}\Delta^{n]a} -
\Delta^{mnas}U^{[i}\Delta^{k]b} - \Delta^{mnbs}U^{[i}\Delta^{k]a}
\right]
\right\}+
$$
$$
+ \frac14\nu_{12} \left\{
-\frac23 U^a \Delta^{ls}\left(U^{[m}
\Delta^{n]bik}+U^{[i}\Delta^{k]bmn} \right)
-\frac23 U^l \Delta^{ab}\left(U^{[m}
\Delta^{n]sik}+U^{[i}\Delta^{k]smn} \right) +
\right.
$$
$$
\left.
+ U^a\left[ U^{[m}\Delta^{n](l}
\Delta^{s)bik}+U^{[i}\Delta^{k](l}\Delta^{s)bmn} -
U^{[m}\eta^{n]b(l} \eta^{s)ik}-
U^{[i}\eta^{k]b(l} \eta^{s)mn}\right]+
\right.
$$
\begin{equation}
\left.
+ U^l\left[ U^{[m}\Delta^{n](a}
\Delta^{b)sik}+U^{[i}\Delta^{k](a}\Delta^{b)smn} -
U^{[m}\eta^{n]s(a} \eta^{b)ik}-
U^{[i}\eta^{k]s(a} \eta^{b)mn}\right]
\right\}
\,.
\label{Y}
\end{equation}


\section*{References}

\begin{thebibliography}{99}

\bibitem{J1} T. Jacobson and D. Mattingly, Phys. Rev. D {\bf 64}
(2001) 024028.

\bibitem{J2} C. Heinicke, P. Baekler and F. W. Hehl, Phys. Rev. D {\bf
72} (2005) 025012.

\bibitem{J3} B. Z.  Foster, Phys.  Rev. D {\bf 73} (2006) 024005.

\bibitem{J41} C. Eling and T. Jacobson, Class. Quant. Grav. {\bf 23}
 (2006) 5625.

\bibitem{J42} C. Eling and T. Jacobson, Class. Quant. Grav.
{\bf 23}  (2006) 5643.

\bibitem{J5} C. Eling, T. Jacobson and M. C. Miller, Phys. Rev. D {\bf
76} (2007) 042003.

\bibitem{J6} T. Jacobson, PoSQG-Ph  (2007) 020.

\bibitem{J7} E. Barausse, T. Jacobson and T. P. Sotiriou, Phys. Rev. D
{\bf 83}  (2011) 124043.

\bibitem{bbm}
P. Berglund, J. Bhattacharyya and David Mattingly,
Phys. Rev. Lett.
{\bf 110} (2013) 071301.

\bibitem{J8} T. Jacobson, arXiv:1310.5115 [gr-qc].

\bibitem{CW} C. M. Will, {\it Theory and experiment in gravitational
physics}, Cambridge University Press, Cambridge, 1993.

\bibitem{Vec0} K. Bamba, S. Nojiri and S.
D. Odintsov, Phys. Rev. D, {\bf 77} (2008) 123532.

\bibitem{Vec1} J. B. Jimenez and A.
L. Maroto, Phys. Rev. D {\bf 80} (2009) 063512.

\bibitem{N1} C. M. Will and K. Nordtvedt, Astrophys. J., {\bf
177} (1972) 757.

\bibitem{N2} K. Nordtvedt and C. M. Will, Astrophys. J., {\bf
177}  (1972) 775.

\bibitem{N3} R. W. Hellings and K. Nordtvedt, Phys. Rev. D {\bf
7} (1973) 3593.

\bibitem{Mo} C. M{\o}ller, The Theory of Relativity, Clarendon
 Press, Oxford, 1952.

\bibitem{LL} L. D. Landau, E. M. Lifshitz, and L. P. Pitaevskii,
{\it Electrodynamics of Continuous Media}, Butterworth Heinemann, Oxford
1960, (second edition, Elsevier Butterworth Heinemann, Oxford,
1984).

\bibitem{ME} A. C. Eringen and G. A.  Maugin, {\it Electrodynamics of
Continua}, Volumes I and II, Springer-Verlag, New York, 1990.

\bibitem{HehlObukhov} F.W. Hehl and Yu. N. Obukhov,
{\it Foundations of classical electrodynamics: Charge, flux, and
metric}, Birkh\"auser, Boston, 2003.

\bibitem{BD09} A. B. Balakin and H. Dehnen, Phys. Lett. B {\bf 681}
 (2009) 113.

\bibitem{LIV31} A. Kostelecky and M. Mewes, Phys. Rev. D {\bf 80}
 (2009) 015020.

\bibitem{LIV2} S. Liberati and L. Maccione,
Ann. Rev. Nucl. Part. Sci. {\bf 59} (2009) 245.

\bibitem{LIV3} A. Kostelecky and N. Russell, Rev. Mod. Phys. {\bf 83}
 (2011) 11.

\bibitem{LIV4} S. Nojiri and S. D. Odintsov, Phys. Rept. {\bf 505} (2011) 59.

\bibitem{LIV1} S. Liberati, Class. Quant. Grav. {\bf 30}
(2013) 133001.

\bibitem{EFT1} C. P. Burgess, Ann. Rev. Nucl. Part. Sci. {\bf 57}
(2007) 329.

\bibitem{EFT2} S. Weinberg, Phys. Rev. D {\bf 77}  (2008) 123541.

\bibitem{EFT3} B. Withers, Class. Quant. Grav. {\bf 26}
(2009) 225009.

\bibitem{EFT4} S. Weinberg, Effective Field Theory, Past and
Future, arXiv:0908.1964 [hep-th].

\bibitem{EFT5} S. Liberati,
Lect. Notes Phys. {\bf 870} (2013) 297.

\bibitem{L1} H. Leutwyler, Annals Phys. {\bf 235} (1994) 165.

\bibitem{L2} T. Becher and H. Leutwyler, Eur. Phys. J. C {\bf 9} (1999)
643.

\bibitem{3} N. Yunes and F. Pretorius, Phys. Rev. D {\bf 80}
(2009) 122003.

\bibitem{2} T. Baker, P. G. Ferreira and C. Skordis, Phys. Rev. D {\bf
87}  (2013) 024015.

\bibitem{wave1} T. Jacobson and D. Mattingly, Phys. Rev. D {\bf 70}
(2004) 024003.

\bibitem{wave2} K. Yagi, D. Blas, E. Barausse and N. Yunes,
Phys. Rev. D {\bf xx}
(2014) xxxxxx; arXiv:1311.7144 [gr-qc].

\bibitem{solbar}
A. R. Solomon and J. D. Barrow, arXiv:1309.4778 [astro-ph.CO].

\bibitem{bumblebee1} R. Bluhm, N.L. Gagne, R. Potting and
A. Vrublevskis, Phys. Rev. D {\bf 77} (2008) 125007.

\bibitem{bumblebee2} M.D. Seifert, Phys. Rev. D {\bf 81} (2010)
065010.

\bibitem{bumblebee3} A. Kostelecky and J. Tasson, Phys. Rev. D {\bf
83} (2011) 016013.

\bibitem{OM1} W. Gordon, Ann. Phys.  {\bf 72} (1923) 421.

\bibitem{OM2} V. Perlick,  {\it Ray Optics, Fermat's Principle, and
Applications to General Relativity}, Springer-Verlag, Berlin (2000).

\bibitem{CM1} A.B. Balakin and W. Zimdahl, Gen. Rel. Grav.  {\bf 37}
(2005) 1731.

\bibitem{CM2} A.B. Balakin, H. Dehnen and A.E. Zayats, Phys. Rev. D
{\bf 76} (2007) 124011.

\bibitem{CM3} A.B. Balakin, H. Dehnen and A.E. Zayats, Annals
Phys. {\bf 323} (2008) 2183.

\bibitem{Fermi} V. Vasileiou et al., arXiv:1305.3463 [astro-ph.HE].

\bibitem{B07} A. B. Balakin, Gravit. Cosmol. {\bf 13} (2007) 163.

\bibitem{BL05} A. B. Balakin and J. P. S. Lemos,
Class. Quant. Grav. {\bf 22} (2005) 1867.

\bibitem{BBL12} A. B. Balakin, V. V. Bochkarev and J. P. S. Lemos,
Phys. Rev. D {\bf 85} (2012) 064015.

\bibitem{BL02} A. B. Balakin and J. P. S. Lemos, Class. Quantum
Grav. {\bf 19} (2002) 4897.

\bibitem{GW1} A. B. Balakin, Class. Quant. Grav. {\bf 14} (1997) 2881.

\bibitem{GW2} A. B. Balakin and
J. P. S. Lemos. Class. Quant. Grav. {\bf 18} (2001) 941.

\bibitem{GW3} A. B. Balakin, R. Kerner and J. P. S. Lemos,
Class. Quant. Grav. {\bf 18} (2001) 2217.

\bibitem{GW4} T. Yu. Alpin and A. B. Balakin, Gravit. Cosmol. {\bf 12}
(2006) 307.

\bibitem{Num2} A. B. Balakin and W.-T. Ni, Class. Quant. Grav. {\bf
27} (2010) 055003.

\bibitem{How} F. W. Hehl and Yu. N. Obukhov, Lect. Notes Phys. {\bf
562} (2001) 479.

\bibitem{DH} I. T. Drummond and S. J. Hathrell, Phys. Rev. D {\bf 22}
 (1980) 343.

\bibitem{BLZ2010} A. B. Balakin, J. P. S. Lemos and A. E. Zayats,
Phys. Rev. D {\bf 81} (2010) 084015.


\end{thebibliography}
\end{document}